\documentstyle[preprint,aps,epsfig]{revtex}  % for preprint form
\tighten       %Gives single-space (RevTex 3.0)

\begin{document}
\draft %prints PACS numbers in

\title{Synchronization in populations of globally
coupled oscillators with inertial effects}
\author{ J.\ A.\ Acebr\'on$^{a,b}$, L.L.
Bonilla$^{a,}$\cite{bonilla:email}, and
R. Spigler.$^{b}$ }
\address{$^{a}$Escuela Polit\'ecnica Superior,  Universidad Carlos III de
Madrid, Avda. Universidad, 20 \\ 28911 Legan{\'e}s, Spain\\
$^{b}$Dipartimento di Matematica, Universit\`{a} di ``Roma Tre",
Largo San Leonardo Murialdo, 1 \\ 00146 Roma, Italy}
\date{\today %September 29, 1999
}

\maketitle
\begin{abstract}
A model for synchronization of globally coupled
phase oscillators including ``inertial'' effects
is analyzed. In such a model, both oscillator
frequencies and phases evolve in time. Stationary
solutions include incoherent (unsynchronized) and
synchronized states of the oscillator
population.  Assuming a Lorentzian distribution
of oscillator natural frequencies, $g(\Omega)$,
both larger inertia or larger frequency spread
stabilize the incoherent solution, thereby making
harder to synchronize the population. In the
limiting case $g(\Omega)=\delta(\Omega)$, the
critical coupling becomes independent of inertia.
A richer phenomenology is found for bimodal
distributions. For instance, inertial effects may
destabilize incoherence, giving rise to
bifurcating synchronized standing wave states.
Inertia tends to harden the bifurcation from
incoherence to synchronized states: at zero
inertia, this bifurcation is supercritical
(soft), but it tends to become subcritical (hard)
as inertia increases. Nonlinear stability is
investigated in the limit of high natural
frequencies. 
\end{abstract}


%\begin{multicols}{2}
%\narrowtext

\section{Introduction}

The dynamical behavior of large populations of
nonlinearly coupled oscillators may describe many
phenomena in Physics, Biology and Medicine,
\cite{strogatz,winfree,arenas}. In particular
synchronization of mean-field coupled phase
oscillators with different natural frequencies is
nicely illustrated by Kuramoto's well-known and
extesively analyzed model
\cite{kuramoto,sakaguchi}. To describe certain
biological phenomena, inertial effects should be
added to this model. In \cite{ermentrout},
Ermentrout revisited the special problem of
self-synchronization in populations of fireflies
of a certain kind ({\it the Pteroptyx malaccae}).
Compared to observed behavior, the approach to
oscillator synchronization as described by the
Kuramoto model seems to be too fast. Thus a more
appropriate adaptive frequency model has been
proposed in \cite{ermentrout,tanaka,acebron2},
where the natural frequency of an oscillator is
a new independent variable, which is allowed to
vary with time. Thus an oscillator is described
by its phase and frequency. From the mathematical
standpoint, the new model is governed by a system
of coupled second-order differential equations
containing inertial terms, in contrast to the
system of first-order differential equations
governing the Kuramoto model. Indeed inertia
slows down synchronization and this may result in
better agreement  between theory and experimental
measurements. Other possible biological
applications of Ermentrout-type models include
after-effects in alterations of circadian cycles
in mammalians, cf.\ \cite{ermentrout}

A different set of applications for oscillators
with inertia include power systems described by
the  swing equations \cite{swing}, or by Hamilton
equations \cite{hamilton}. An important
technologically relevant application is the study
of superconducting Josephson junctions arrays
\cite{wiesenfeld,park}. Here inertial terms
describe the effect of nonzero electrical
capacitance. Such effect is often far from
being negligible, and it is absent in the Kuramoto
model, \cite{vanduzer}.

In this paper, we consider the model equations of 
Ref.\ \cite{acebron2},
\begin{eqnarray}
 \dot{\theta_{j}} &=&\omega_{j}\nonumber\\
m\,\dot{\omega_{j}} &=& -\omega_{j}+\Omega_{j}+
K\,r_{N} \sin(\psi_{N} -\theta_{j}) +
\xi_{j}(t),\label{1} \\ 
&&\quad \quad \quad \quad
\quad \quad \quad \quad \quad \quad \quad
j=1,\ldots,N,   \nonumber
\end{eqnarray}
where $\theta_{j}$, $\omega_{j}$ and $\Omega_{j}$
denote phase, frequency and natural frequency of
the $j$th oscillator, respectively. The natural
frequencies are distributed with probability
density $g(\Omega)$, which may have a single
maximum ({\em unimodal} distribution), or several
peaks ({\em multimodal} distribution). The
positive parameters $m$ and $K$ are the
``inertia'' and the coupling strength,
respectively. The complex order parameter defined
by
\begin{equation} r_{N}
e^{i\,\psi_{N}}=\frac{1}{N} \sum_{j=1}^{N} 
e^{i\,\theta_{j}},\label{2}
\end{equation}
measures phase synchronization: $r_N>0$ if the
oscillators are synchronized and $r_N=0$ if not.
Finally, $\xi_{j}$'s are independent identically 
distributed Gaussian white noises, with $\langle
\xi_{j} \rangle =0,\,\langle\xi_{i}(t) \xi_{j}(s)
\rangle = 2D \delta_{ij} \delta(t-s)$. White noise
terms were not included in
\cite{ermentrout,tanaka}. When the inertial terms
vanish, $m=0$, Eqs.\ (\ref{1}) and (\ref{2})
are exactly the Kuramoto model.

Typically, {\it N} is very large, and oscillator
synchronization is conveniently analyzed in the
limiting case of infinitely many oscillators. In
this limit, models with mean-field coupling are
described by an evolution equation for the
one-oscillator probability density,
$\rho(\theta,\omega,\Omega,t)$, \cite{bon87}. For
the present model this equation is
\cite{acebron2}
\begin{eqnarray}
\frac{\partial \rho}{\partial t}&=&\frac{D}{m^2}
\frac{\partial^{2} \rho}{\partial
\omega^{2}} \nonumber\\
&-&\frac{1}{m}\frac{\partial}{\partial \omega}
[\big(-\omega+\Omega+K r
\sin(\psi-\theta)\big)\rho]
-\omega\frac{\partial \rho}{\partial\theta}\,,
\label{fpe}
\end{eqnarray}
where the order parameter is now given by
\begin{eqnarray}
r e^{i \psi}= \int_{0}^{2\pi}
\int_{-\infty}^{+\infty}
\int_{-\infty}^{+\infty}  e^{i \theta}
\rho(\theta,\omega,\Omega,t)
g(\Omega)\,d\Omega\, d\omega\, d\theta.
\label{order}
\end{eqnarray}
Equations (\ref{fpe}) and (\ref{order}) should be 
supplemented with appropriate initial and
boundary data ($\rho$ is $2\pi$-periodic in
$\theta$ and has suitable decay behavior as
$\omega\to\pm\infty$) plus the normalization
condition,
\begin{equation}
\int_0^{2\pi} \int_{-\infty}^{+\infty}
\rho(\theta,\omega, \Omega,t)\, d\omega\,
d\theta=1. 
\label{norma}
\end{equation}
Differentiating $\int_0^{2\pi} \int_{-\infty}^{
+\infty} \rho(\theta, \omega,\Omega,t)\,
d\omega\,d\theta$ with respect to time, and then
using Eq.\ (\ref{fpe}) itself, together with
periodicity in $\theta$ and decay in $\omega$, we
find that the left side of (\ref{norma}) is time
independent. Normalization to unity of the
initial probability density then implies
(\ref{norma}) for the solution of  (\ref{fpe}). 

In this paper, we study oscillator synchronization
and transition from incoherence to synchronization
in the model (\ref{fpe}) - (\ref{norma}). The 
incoherent solution of (\ref{fpe}) - (\ref{norma})
(or simply {\em incoherence}) is a stationary
solution which is independent of $\theta$. This
solution asigns equal probability to all angles
and has $r=0$ (no order), so it corresponds to
lack of oscillator synchronization. There are
synchronized solutions which branch off from
incoherence as the coupling among oscillators is
increased. These bifurcations describe the
synchronization transitions, which we have
analyzed and compared to the corresponding ones
in the Kuramoto model. Our main results are that
inertia: (i) may stabilize incoherence, making it
harder to synchronize oscillators, and (ii) it
may harden the synchronization transition. In the
Kuramoto model ($m=0$) or with oscillators with
identical natural frequencies, the
synchronization transition is soft (supercritical
bifurcation), whereas it may become hard
(subcritical bifurcation) if the distribution of
natural frequencies has a nonzero spread
(unimodal Lorentzian distribution) or several
peaks (e.g., a discrete bimodal distribution).
The methods we have used in our analysis are
similar to those previously employed in the
Kuramoto model
\cite{strogatz2,bns,crawford,bps,acebron1}:
linear stability of incoherence, bifurcation
analysis, high-frequency singular perturbations
and numerical solutions. An important difference
is that now we do not have an explicit functional
form for stationary solutions (as it was the case
for the Kuramoto model). This has led us to use
mode-coupling expansions of the solution and
solving the corresponding mode-coupling
equations. Solutions of these equations in close
form are not always accesible, so that we have
introduced some closure assumptions. The results
of these uncontrolled assumptions have been
compared to direct simulations or to approximate
amplitude  equations and found reasonable in the
limit of small inertia. 

The rest of the paper is as follows. In Section
II, we find the incoherent solution and study its
linear stability for several natural frequency
distributions. Results are compared with those
obtained in the massless case
\cite{strogatz2,bns,crawford,bps,acebron1}. It is
found that the critical coupling needed to
destabilize incoherence increases with $m$ for
``unimodal'' frequency  distributions of the
Lorentzian type. The critical coupling is
independent of $m$ when $g(\Omega) =
\delta(\Omega)$. In this case the time needed to
reach synchronization increases as $m$ increases.
If $g(\Omega)= [\delta(\Omega-\Omega_0) + \delta(
\Omega +\Omega_0)]/2$ (discrete bimodal
distribution), the critical coupling may grow or
decrease with $m$ depending on the values of
$\Omega_0$. In Section III, we construct other 
stationary solutions by two procedures: an
amplitude expansion for solutions branching off
from incoherence and a general expansion in
Hermite polynomials which is appropriately
truncated. An exact analytical solution is
obtained if $g(\Omega) = \delta(\Omega)$ (cf.
\cite{acebron2}), while analytical
approximations for small $m$ and $\Omega$ are
available in the general case. We have observed
that inertia tends to {\em harden} the
synchronization transition: in the Kuramoto model
($m=0$) or with oscillators with identical
natural frequencies, the synchronization
transition is soft (supercritical bifurcation),
whereas it becomes hard (subcritical bifurcation)
in the cases of unimodal Lorentzian or discrete
bimodal frequency distributions. In Section IV, we
obtain approximations to stable time-dependent
solutions of Eq.\ (\ref{fpe}) in the ``high
frequency limit'', $\Omega\to\infty$
\cite{acebron1}. There are partially synchronized
nonlinearly stable solutions of standing wave
type, as in the Kuramoto model
\cite{crawford,bps}). Finally, numerical results
are presented in Section V, and compared to the
approximate or exact solutions of previous
Sections. Two Appendices at the end are devoted to
technical details. 

\section{Linear stability of the incoherent
solution}
The incoherent solution is a
$\theta$-independent stationary solution of
(\ref{fpe}). Its order parameter is $r=0$
according to (\ref{order}). Then Eqs.\
(\ref{fpe}), decay as $\omega\to\pm\infty$ and
the normalization condition (\ref{norma}) yield
the incoherent solution:
\begin{equation}
\rho_{0}(\omega,\Omega)=\frac{1}{2\pi} \sqrt{
\frac{m}{2\pi D}} e^{-\frac{m}{2D}(
\omega-\Omega)^{2}}. \label{incoherent}
\end{equation}
To analyze its linear stability, let us consider a small disturbance
about incoherence,
\begin{equation}
\rho(\theta,\omega,\Omega,t)=\rho_{0}(\omega,\Omega)+ \varepsilon\,
\eta(\theta,\omega,\Omega,t)+ O(\varepsilon^2),\label{perturb}
\end{equation}
where $\varepsilon\ll 1$. Normalization of
$\rho(\theta,\omega,\Omega,t)$ then implies
\begin{equation}
\int_0^{2\pi}\int_{-\infty}^{+\infty}
\,\eta(\theta,\omega,\Omega,t)\,d\omega\,d\theta=0.\label{zero}
\end{equation}
We now introduce (\ref{perturb}) into (\ref{fpe}) and (\ref{order})
and equate like terms in $\varepsilon$. To order $\varepsilon$, the
result is
\begin{eqnarray}
\frac{\partial \eta}{\partial t} + \omega\frac{\partial \eta}
{\partial \theta} - \frac{1}{m}\frac{\partial}{\partial  \omega}
[(\omega-\Omega)\,\eta] - \frac{D}{m^2}\frac{\partial^2 \eta}{\partial
\omega^2} = \nonumber\\
-\frac{K\, \frac{\partial\rho_{0}}{\partial\omega}}{m}\,
\int_0^{2\pi}\int_{-\infty}^{+\infty}\int_{-\infty}^{+\infty}
\eta(\phi,\omega,\Omega,t)\, \nonumber\\
\times \sin(\phi-\theta)\, g(\Omega) d\Omega d\omega d\phi.
\label{eta}
\end{eqnarray}
We now insert a trial solution
\begin{equation}
\eta(\theta,\omega,\Omega,t) = e^{\lambda\,t}\,
\sum_{n=-\infty}^{\infty} b_n(\omega;\Omega,\lambda)\,
e^{i n \theta} \label{trial}
\end{equation}
(which is $2\pi$-periodic in $\theta$) into (\ref{eta}), thereby
obtaining
\begin{eqnarray}
\frac{d^2 b_{n}}{d\omega^2} + \frac{m\, (\omega-\Omega)}{D}\,
\frac{db_{n}}{d\omega} + {m\, (1-m\lambda-inm\omega)\,
b_{n}\over D} =  \nonumber\\
 {\pi m K \, (i \delta_{n,1} - i \delta_{n,-1})\,
{\partial\rho_{0}\over \partial \omega} \over D}\,  \langle 1,b_n
\rangle ,\label{eqb} \end{eqnarray}
where we have defined the scalar product
\begin{eqnarray}
\langle \varphi,\psi \rangle = \int_{-\infty}^{+\infty}
\int_{-\infty}^{+\infty} \overline{\varphi(\omega,\Omega)}
\psi(\omega,\Omega)\, g(\Omega)\, d\Omega\, d\omega.
\label{scalarprod}
\end{eqnarray}
Notice that $b_{-n} = \overline{b_{n}}$ and that $b_n = 0$ because of
the normalization condition (\ref{zero}).

Equation (\ref{eqb}) can be transformed into a nonhomogeneous
parabolic cylinder equation by the following change of variable:
\begin{eqnarray}
b_{n}(\omega;\Omega,\lambda) = \exp\left[- {m\, (\omega - \Omega)^{2}
\over 4D} \right]\, \beta_n (w;\Omega,\lambda), \label{beta1}\\
w = \sqrt{{m\over D}}\, (\omega - \Omega + 2nDi) .
\label{beta2}
\end{eqnarray}
Inserting (\ref{beta1}) and (\ref{beta2}) into (\ref{eqb}), we
obtain
\begin{eqnarray}
\frac{d^{2} \beta_{n}}{dw^2}+ \left[\frac{1}{2} - {w^{2}\over 4}
- m\, (\lambda + in\Omega + n^2 D)\right]\, \beta_n = \quad
\nonumber\\
i\pi K {\partial\rho_{0}\over \partial \omega}
e^{{1\over 4}(w-2i\sqrt{mD})^{2}}\, \langle 1,
e^{-{1\over 4}(w-2i\sqrt{mD})^{2}} \beta_1 \rangle\,
\delta_{n,1}.  \label{beta3}
\end{eqnarray}
(Recall that $d\omega = \sqrt{D/m}\, dw$ when using the definition
of scalar product). Let us assume now that $n\neq\pm 1$ and that
$\Omega$ is a fixed real number. Then the right hand side of
(\ref{beta3}) is zero and the resulting equation has the following
eigenvalues
\begin{eqnarray}
\lambda_{p,n}(\Omega)
= -{p\over m} - n^2 D - i n \Omega,\quad p = 0, 1, 2,\ldots ,
\label{beta4}
\end{eqnarray}
associated to the eigenfunctions
\begin{eqnarray}
\beta_{p,n}(w;\Omega,\lambda_{p,n}) = D_p (w) = 2^{-{p\over 2}}\,
e^{-{w^{2}\over 4}} \, H_p\left({w\over\sqrt{2}}\right) ,
\label{beta5}
\end{eqnarray}
which are independent of $n$ and $\Omega$. In this formula, $D_p(w)$
and $H_p(x)$ are the parabolic cylinder function and the Hermite
polynomial of index $p$, respectively \cite{gra,abramovitz}. The
eigenvalues $\lambda_{p,n}(\Omega)$ of (\ref{beta4}), with $n=\pm 1,
\pm 2, \ldots$, $p=0,1, \ldots$ and $\Omega$ belonging to the support
of  $g(\Omega)$, constitute the continuous spectrum of the linear
stability problem. In fact, a nonhomogeneous linear problem with a
homogeneous part given by (\ref{beta3}) cannot be solved for an
arbitrary source term if $\lambda = \lambda_{p,n}$. Notice
that the continuous spectrum lies to the left side of the imaginary
axis if $D>0$ and $n\neq 0$. Then the ``eigenvalues'' (\ref{beta4})
have negative real parts (and therefore correspond to stable modes).
As we have already said, the neutrally stable modes with $n=0$ have
zero amplitude due to the normalization condition (\ref{zero}).

If $n=1$, we can solve (\ref{eqb}) by means of an expansion
in eigenfunctions $D_p(w)$, $p = 0,1,2,\ldots$. To obtain the
generalized Fourier coefficients of $\beta_1$, we multiply both
sides of (\ref{eqb}) by $D_p(w)$ and integrate over $w$. As
$\int_{-\infty}^{\infty} D_p(x) D_n(x) dx = \sqrt{2\pi}\, p!\,
\delta_{pn}$ (orthogonality condition, cf.\ \S 7.711.1 of
\cite{gra}), the result is
\begin{eqnarray}
\beta_1(\omega;\Omega,\lambda) &=& -{i\pi K\over m}\,
\langle 1,e^{-\left({w\over 2} - i\sqrt{mD}\right)^{2}}
\beta_1 \rangle  \nonumber\\
&\times & \sum_{p=0}^{\infty} {\int_{-\infty}^{\infty}
e^{\left({w\over 2} - i\sqrt{mD}\right)^{2}} D_p
\rho'_{0} dw\over \sqrt{2\pi} p! \left({p\over m} + \lambda
+ i\Omega + D\right)}\, D_p(w), \label{beta6}
\end{eqnarray}
where
\begin{eqnarray}
\rho'_{0}(w) &=& \left. {\partial\rho_{0}\over\partial\omega}
\right|_{\omega= \Omega - i2D + (D/m)^{{1\over 2}} w} \nonumber\\
&=& - {m\, (w - i2\sqrt{mD})\over (2\pi)^{{3\over 2}} D}\,
e^{-{1\over 2}(w - i2\sqrt{mD})^{2}}\,.
\label{beta7}
\end{eqnarray}

Once we have found $\beta_1$, we can calculate the scalar product
$\langle 1,e^{-\left({w\over 2} - i\sqrt{mD}\right)^{2}}
\beta_1 \rangle$. Since this scalar product appears as a factor in
both sides of the resulting expression, we can divide by it, thereby
obtaining an eigenvalue equation for $\lambda$:
\begin{eqnarray}
1 = \frac{-i \pi K\sqrt{D}}{\sqrt{2\pi m^{3}}}\,
\sum_{p=0}^{\infty} {1\over p!}\,\int_{-\infty}^{+\infty}
e^{\left({w\over 2}  - i\sqrt{mD}\right)^{2}} D_p
\rho'_{0} dw\nonumber\\
\times \int_{-\infty}^{+\infty}
e^{-\left({w\over 2} - i\sqrt{mD}\right)^{2}} D_{p}\, dw
\nonumber\\
\times \int_{-\infty}^{+\infty} {g(\Omega)\over
{p\over m} + \lambda + i\Omega + D}\, d\Omega .
\label{beta8}
\end{eqnarray}
In Appendix A, we show that this equation may be rewritten as
\begin{eqnarray}
1 = \frac{K}{4\pi \sqrt{mD}}\,
\sum_{p=0}^{\infty} { {\cal A}_{p}(\sqrt{mD})\, {\cal
A}'_{p}(\sqrt{mD})\over p!}\,
\nonumber\\
\times \int_{-\infty}^{+\infty} {g(\Omega)\over
{p\over m} + \lambda + i\Omega + D}\, d\Omega ,
\label{beta9}
\end{eqnarray}
where ${\cal A}_p(x)$ is defined as
\begin{eqnarray}
{\cal A}_p(x) = \int_{-\infty}^{+\infty} D_p(w)\,
e^{-\left({w\over 2} - ix\right)^{2}}\, dw .
\label{beta10}
\end{eqnarray}
The result of evaluating this integral is (cf.\ Appendix A):
\begin{eqnarray}
{\cal A}_p(x) = i^{p} \sqrt{2\pi}\, e^{{x^{2}\over 2}}
x^{p} \quad\quad (x>0). \label{beta11}
\end{eqnarray}
Inserting (\ref{beta11}) in (\ref{beta9}), we obtain
\begin{eqnarray}
1 = \frac{K\, e^{m D}}{2}\,\sum_{p=0}^{\infty}
{(-mD)^{p}\,\left(1+ {p\over mD}\right)\over p!}\nonumber\\
\times \int_{-\infty}^{\infty} {g(\Omega)\, d\Omega\over
\lambda+D+i\Omega+\frac{p}{m}}\,.
\label{dispersion3}
\end{eqnarray}
As $m\to 0$, this equation coincides with that obtained for the
Kuramoto model, \cite{strogatz2}. Equation (\ref{dispersion3}) can be
rewritten in terms of incomplete gamma functions as follows
\cite{abramovitz,temme}:
\begin{eqnarray}
\frac{2D}{K} = e^{mD} \int_{-\infty}^{\infty}
\left[mD\,\gamma(m\, (\lambda+D+i\Omega),mD) \right.\nonumber\\
\left. - \gamma(1+(\lambda+D+i\Omega)m,mD) \right]
{g(\Omega) d\Omega\over (mD)^{m(\lambda+D+i\Omega)}} \nonumber\\
= \int_{-\infty}^{\infty}
\left[1 - {(\lambda + i\Omega)m e^{m D} \gamma
(m(\lambda+D+i\Omega),mD)\over (mD)^{(\lambda+D+i\Omega)m}}
\right] \nonumber\\
\times g(\Omega)\, d\Omega\,, \label{gammaf}
\end{eqnarray}
where \cite{abramovitz,temme}
\begin{eqnarray}
\gamma(a,x)=\int_{0}^{x}\,e^{-t} t^{a-1}\, dt.
\end{eqnarray}

>From now on, we analyze Eq.\ (\ref{dispersion3}) for special
frequency distributions:\\
\bigskip

\noindent
(a) {\it Unimodal frequency distribution},
$g(\Omega)=\delta(\Omega)$.

\begin{figure}
\centerline{\hbox{\psfig{figure=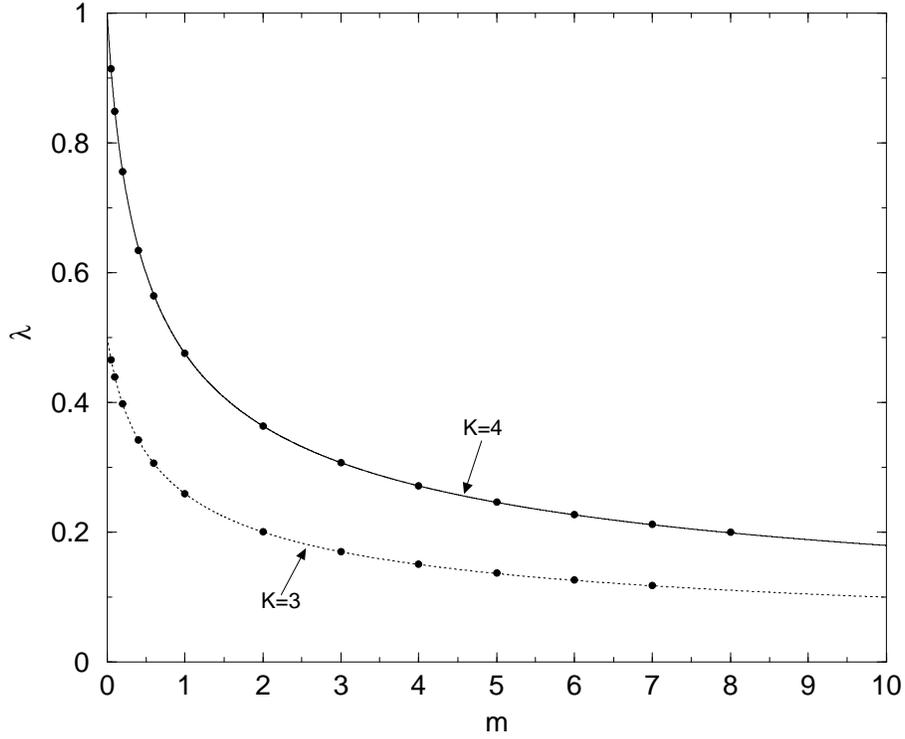,width=12.0cm}}}
\caption{Discrete unimodal frequency distribution:
Comparison between analytical (continuous line),
and numerical (dots) evaluation of the eigenvalue
$\lambda$ as a function of $m$ for $D=1$ and two different
values of the coupling strength $K$.}
\label{lsu1}
\end{figure}

\begin{figure}
\centerline{\hbox{\psfig{figure=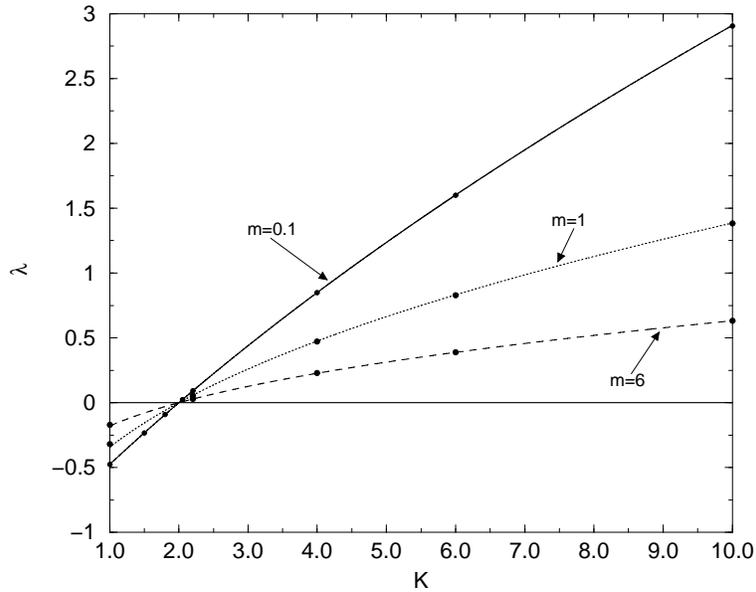,width=10.0cm}}}
\caption{Discrete unimodal frequency
distribution: Eigenvalue $\lambda$ as a function
of $K$ for three different masses $m$. Other
parameter values and meaning of lines as in
Figure 1. Note that curves corresponding to
different masses all intersect the axis $\lambda=0$ at the 
same point, as expected from theory.}
\label{lsu2}
\end{figure}

In this case we show that if $\mbox{Re}\lambda=0$,
then $\mbox{Im}\lambda=0$. Thus, the eigenvalues
that may acquire a positive real part are real.
Then the critical coupling is obtained by
setting $\lambda=0$.  By subtracting from Eq.\
(\ref{dispersion3}) its complex conjugate, we
obtain
\begin{eqnarray}
0 = \mbox{Im}(\lambda)f(\mbox{Im}\lambda,m,D),
\label{zeros}
\end{eqnarray}
where
\begin{equation}
f(\mbox{Im}\lambda,m,D)=\sum_{p=0}^{\infty}
{(-mD)^{p}\,\left(1+ {p\over mD}\right)\over
p!\,\left[(\mbox{Im}\lambda)^2+\left(D+
\frac{p}{m}\right)^{2}\right]}.
\end{equation}
Notice than the even function $f(\mbox{Im}
\lambda,m,D)$ decreases monotonically  with
Im$(\lambda)>0$. On the other hand, $f$ tends to
zero, as $\mbox{Im}\lambda\to +\infty$, and
\begin{equation}
f\sim \frac{m}{D}(m\,D)^{-m\,D}\gamma(m\,D,m\,D)>
0, \quad\quad\mbox{as }\quad
\mbox{Im}\lambda\to 0.
\end{equation}
Thus $f$ does not vanish at finite values of
Im$\lambda$, and therefore the
only solution of (\ref{zeros}) is Im$\lambda=0$.
Setting now $\lambda=0$, Equation
(\ref{dispersion3}) yields the critical coupling
$K=K_c$,
\begin{eqnarray}
\frac{K_c}{2D}\, e^{m\,D}\left[\sum_{n=0}^{\infty}
\frac{(-m\,D)^n}{n!}\right] = 1, \quad \mbox{and
therefore}
\nonumber\\
K_c=2D.\label{unimodalKc}
\end{eqnarray}
Figures \ref{lsu1} and \ref{lsu2} show the largest eigenvalue
$\lambda$ as a function of $m$ and $K$. To compute $\lambda$ numerically,
we used from Eq.  (\ref{perturb}), and (\ref{trial}), that the 
amplitude of the order parameter is
\begin{equation}
r\approx C\,e^{\lambda t},
\end{equation}
close to incoherence. Then, the goal is to simulate the evolution 
of the system, choosing the initial condition sufficiently close to
the incoherent solution, and obtain numerically the amplitude order
parameter $r(t)$. Fig. \ref{lsu2} shows
that different eigenvalue curves (for different
$m$) intersect the horizontal axis, $\lambda=0$,
{\em at the same value of} $K$, as expected from
Eq. (\ref{unimodalKc}).

\noindent
(b) {\it Unimodal Lorentzian frequency
distribution},
$g(\Omega)=\frac{\varepsilon/\pi}{\varepsilon^2+\Omega^2}$.

Eq.\ (\ref{dispersion3}) becomes
\begin{eqnarray}
1=\frac{K}{2D}e^{m
D}\left[\frac{D}{\lambda+D+\varepsilon}+\sum_{n=1}^{\infty}
\frac{(-m\,D)^{n}}{n!} \right. \nonumber\\
\left.\times \frac{n+m\,D}{m(\lambda+D+
\varepsilon)+n}\right]
= \frac{K}{2}m\,e^{mD} \nonumber\\
\times (m\,D)^{-m(\lambda+D+\varepsilon)} \left[(m\,D)^{-m(\lambda+D+
\varepsilon)-1}\, e^{-m\,D} \right. \nonumber\\
\left.
-\gamma (m(\lambda+D+\varepsilon),m\,D)\left(\frac{\lambda+D+
\varepsilon}{D}-1\right) \right]
\label{lorentz}
\end{eqnarray}

An explicit solution for $\lambda$ cannot in general be found. Thus,
we consider several limiting cases corresponding to physically
interesting parameter choices. In the small noise limit, $D\ll 1$, we
consider the cases $(i)$ $m=O(1)$ fixed, and $(ii)$ $mD=1$. It is
remarkable that the expansion
\begin{equation}
\gamma(a,x)=e^{-x} x^{a}\sum_{n=0}^{\infty} \frac{x^n}{(a)_{n+1}},
\label{tt}
\end{equation}
where $(a)_k=a(a+1)\cdots(a+k-1)$, $k=1, 2,
\ldots$, holds in both cases. If $x=m D\to 0$, $a=m\,D+m(\lambda+
\varepsilon)>x$, (\ref{tt}) holds as a convergent expansion,
\cite{temme}. If $x=mD=1$, and $a=1+m(\lambda+\varepsilon)\to\infty$
(with fixed $\lambda$ and $\varepsilon$ of order 1), (\ref{tt}) holds
as an asymptotic expansion, \cite{tricomi}. Inserting (\ref{tt}) in
Eq.\ (\ref{lorentz}), we obtain
\begin{eqnarray}
\frac{2D}{K} = 1+\frac{x-a}{a}\, \left[1+\frac{x}{a+1}+
\frac{x^2}{(a+1)(a+2)}
+ O\left({x^{3}\over a^{3}}\right)\right]
\label{lorentz1}
\end{eqnarray}
Similarly to the unimodal case, it is possible to prove
that $\lambda$ is always real. To this purpose, notice that
replacing $\lambda+\varepsilon$ in eq. (\ref{lorentz})
with $\lambda$ in eq. (\ref{dispersion3}) (setting $\Omega=0$) we obtain
the same equation.
The critical coupling $K=K_c$ is then found by
setting  $\lambda=0$ in (\ref{lorentz1}). In case $(i)$, we have
\begin{equation}
K_c=2 \varepsilon(m\,\varepsilon+1)+\frac{2(2+3m\,\varepsilon)}{2+
m\,\varepsilon}\,D +O\left(D^2\right),
\label{kc1}
\end{equation}
In the limit of vanishing mass, we recover the result $K_c=
2(D+\varepsilon)$ valid for the Kuramoto model \cite{strogatz2}.
Another important limit is $\varepsilon=0$, which reproduces the
unimodal distribution. We find $K_c=2D$, {\it independently} of
mass. Thus the spread in frequency distribution plays an important
role in synchronizing populations of oscillators affected by inertia.

In case $(ii)$, Eq.\ (\ref{lorentz1}) yields
\begin{equation}
1=\frac{K\, x}{2D\, a(a+1)} + O\left(a^{-3}\right),
\end{equation}
from which
\begin{equation}
\lambda=-\left[\varepsilon+\frac{1}{2m}\left(3\pm\sqrt{1+2K\,m}\right)\right].
\end{equation}
This quantity is always real and vanishes for
$K=K_c=2\varepsilon(m\, \varepsilon+3)
+\frac{4}{m}$. Note that $K_c$ grows roughly
linearly with $m$. Thus oscillator
synchronization is made harder by increasing
inertia in the limit of vanishing noise. This
behavior is slightly different from that
described in \cite{tanaka}. There numerical
simulations seemed to show that incoherence
remains stable up to a critical coupling, which
was independent of $m$. The singular nature of
the limit $D\to 0$ makes the cause of this
discrepancy unclear, although we should mention
that no stability analysis was conducted in
\cite{tanaka}. In the opposite limit $m\to\infty$,
$\lambda\to -\varepsilon$, and incoherence is always stable.

\begin{figure}
\centerline{\hbox{\psfig{figure=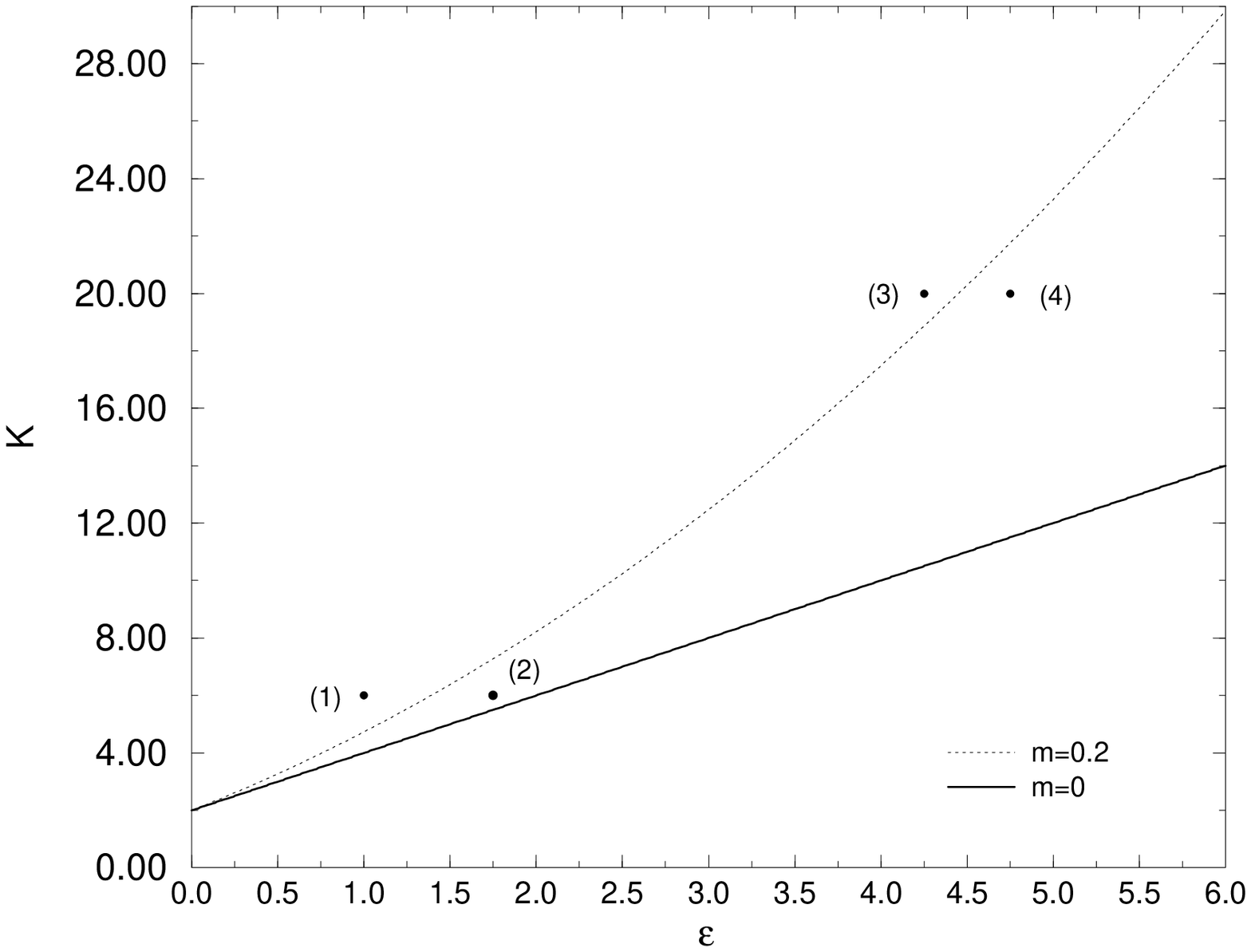,width=12.0cm}}}
\centerline{\hbox{\psfig{figure=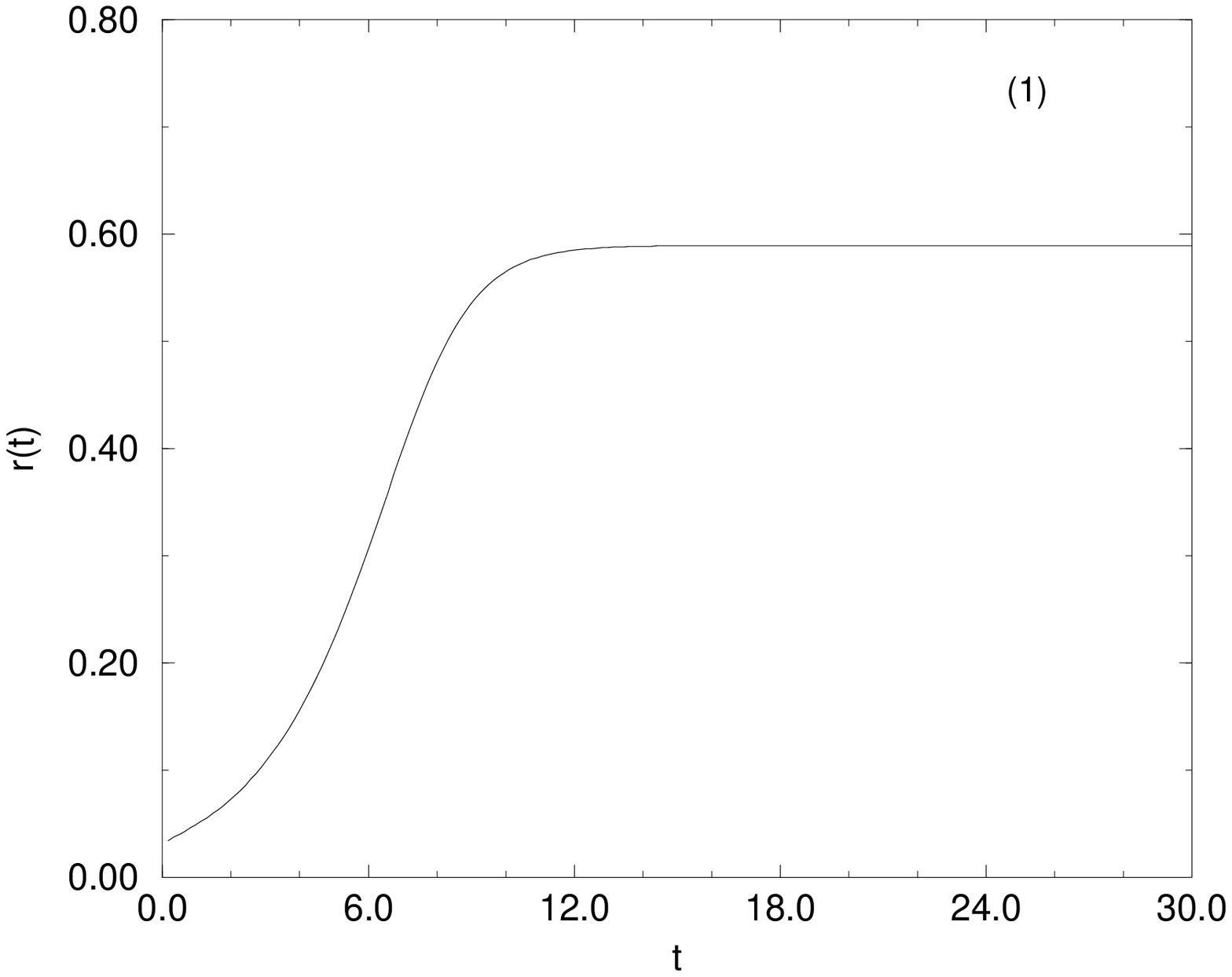,width=7.0cm}}
\hbox{\psfig{figure=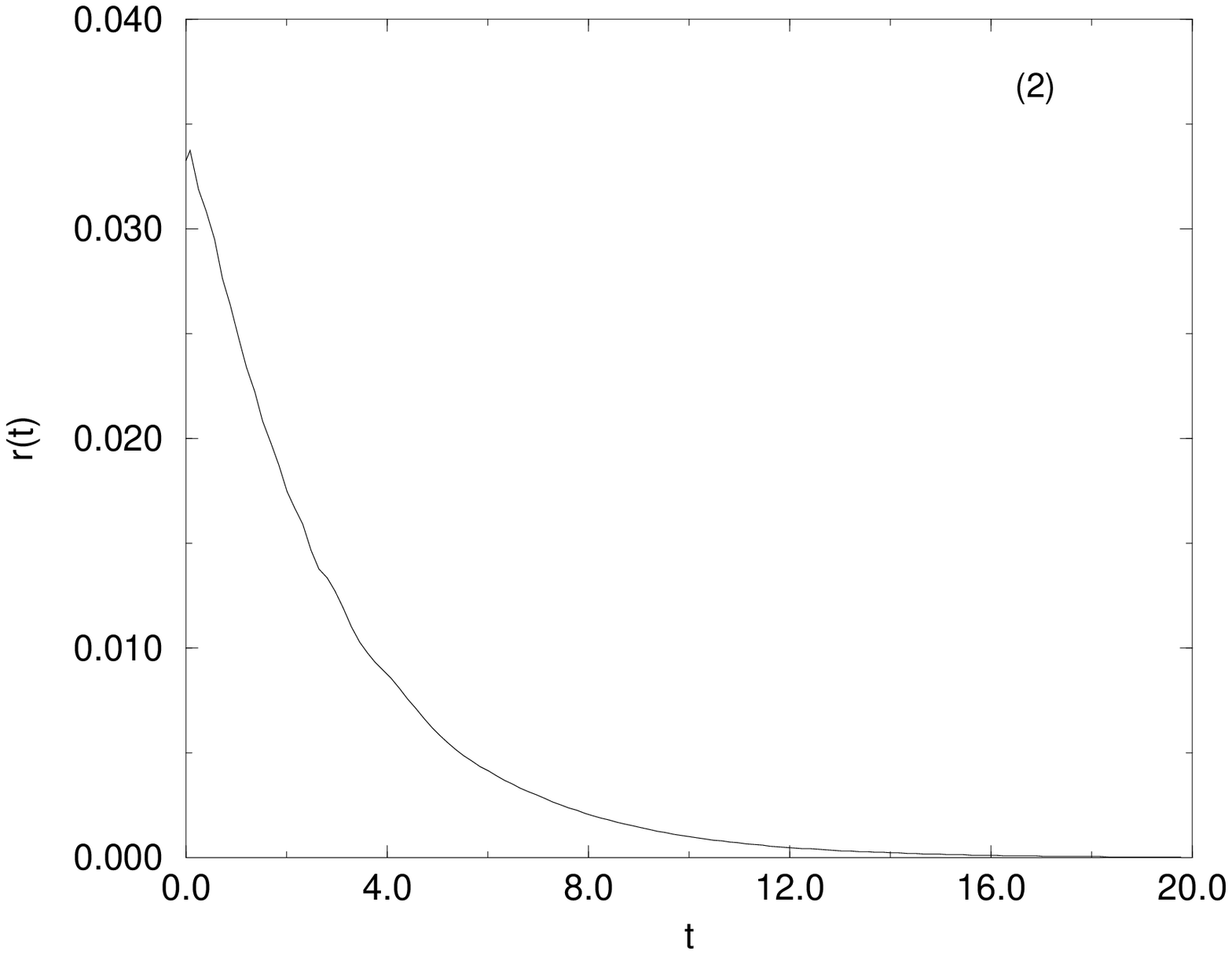,width=7.0cm}}}
\centerline{\hbox{\psfig{figure=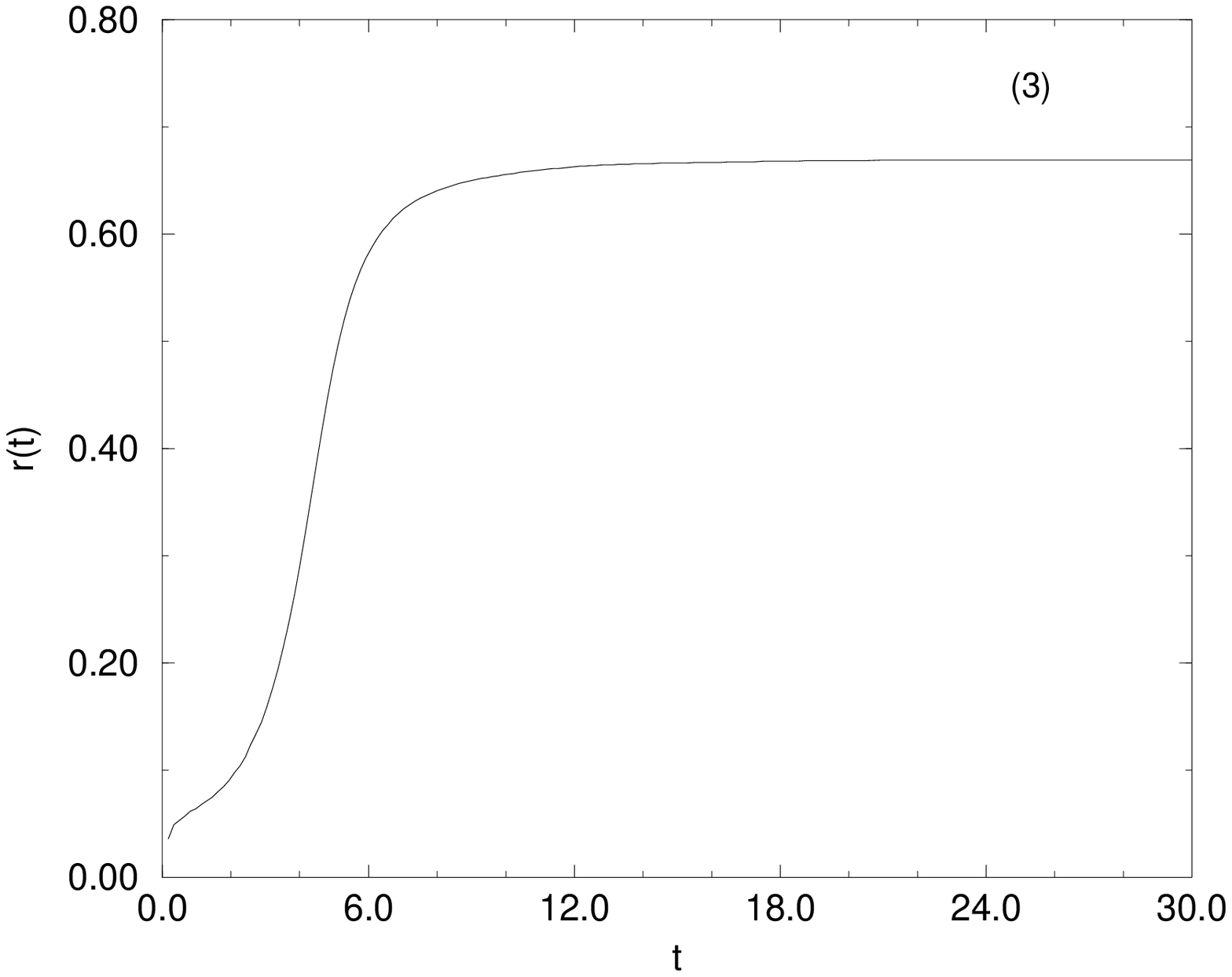,width=7.0cm}}
\hbox{\psfig{figure=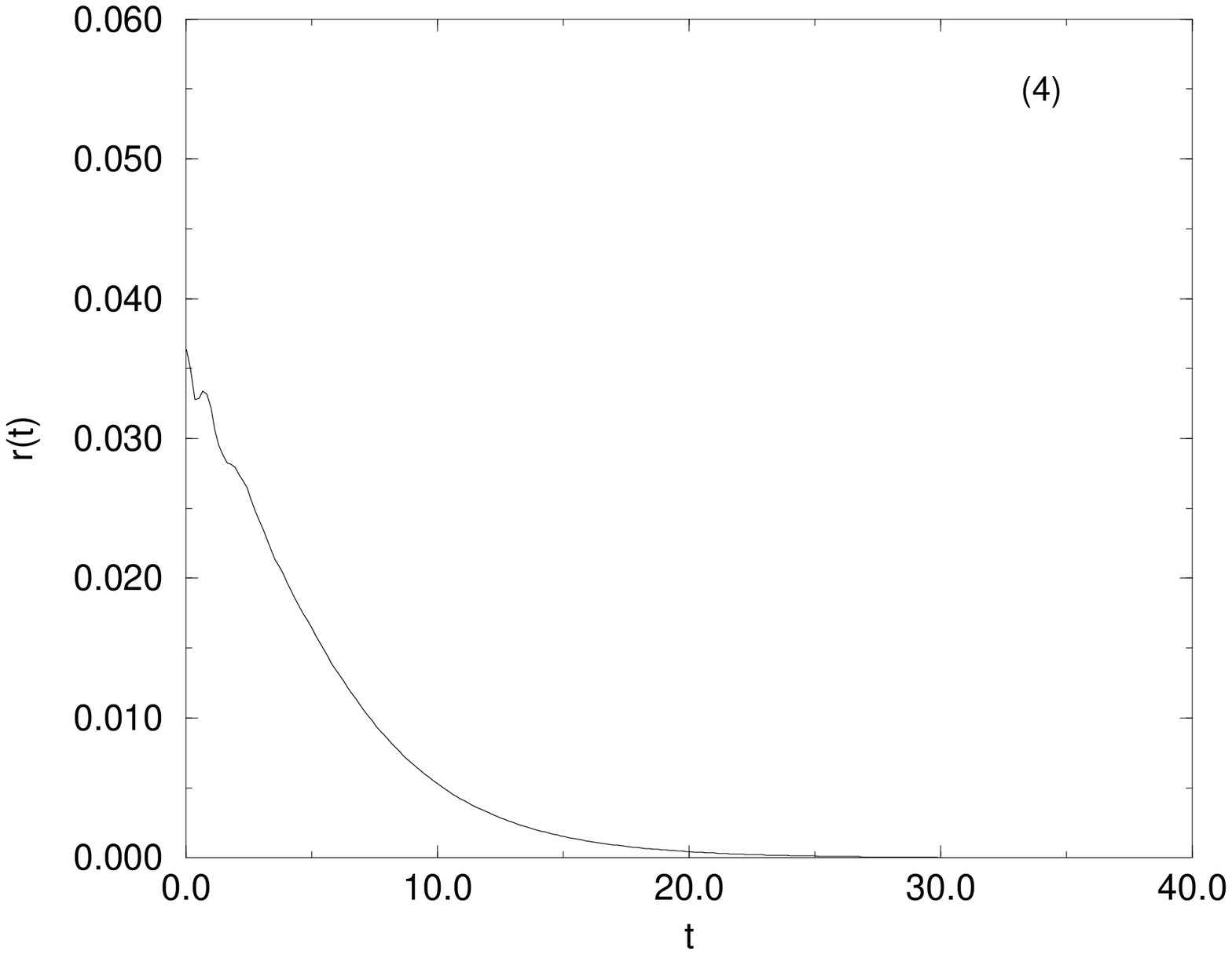,width=7.0cm}}}
\caption{Unimodal Lorentzian frequency
distribution: Stability diagram of incoherence in
the parameter space ($\varepsilon$,$K$) for
$m=0.2$, $D=1$. The amplitude of the order
parameter as a function of time is numerically
calculated and displayed at the points marked by:
(1) $K=6$, $\varepsilon=1$; (2) $K=6$,
$\varepsilon = 1.75$; (3) $K=20$, $\varepsilon
=4.25$; (4) $K=20$, $\varepsilon=4.75$. }
\label{lsl1}
\end{figure}

 The stability diagram in
the parameter space ($\varepsilon$,$K$) is
shown in Figure \ref{lsl1} for 
$m=0.2$, and compared to that of the
Kuramoto model ($m=0$). This diagram is obtained 
 from Eq. (\ref{lorentz}) with $\lambda=0$, for
fixed $D$ and $m$. In this figure, we have also
plotted the  evolution of the order parameter
amplitude for the parameter values marked in the
stability diagram by (1) to (4). In all cases,
the initial condition is taken sufficiently  close
to the incoherent solution, $r=0$.
\bigskip

\noindent
(c) {\it Bimodal frequency distribution},
$g(\Omega)=\frac{1}{2}[\delta(\Omega-\Omega_0)
+\delta(\Omega+\Omega_0)]$.

Eq.\ (\ref{dispersion3}) becomes
\begin{eqnarray}
\frac{K\, e^{m D}}{2D}\,\sum_{n=0}^{\infty}
\frac{(-m\,D)^{n}}{n!}\,\frac{(n+m\,D)[m(\lambda+D)+n]}
{[m(\lambda+D)+n]^2+m^2\Omega_0^2} = 1.\label{beta12}
\end{eqnarray}
In the high frequency limit, $\Omega_0\to\infty$, we can find an
analytical formula for $\lambda$ by inserting the following asymptotic
expansion for the incomplete gamma function in (\ref{gammaf})
\cite{tricomi},
\begin{equation}
\gamma(a,x)\approx \frac{e^{-x}x^{a}}{a},\quad\quad a\to\infty,
\end{equation}
where $a=m(\lambda+D+i\,\Omega_0)$, and $x=m\,D$. The result is
\begin{equation}
1=\frac{K}{4}\left(\frac{1}{\lambda+D+i\,\Omega_0}+\frac{1}{\lambda
+D-i\,\Omega_0} \right),
\end{equation}
which yields
\begin{equation}
\lambda=-D+\frac{K}{4}+\frac{K}{4}i\,\sqrt{16\Omega_{0}^2-K^2}.
\end{equation}
Re$\lambda = 0$ gives the same critical coupling as the Kuramoto model,
$K_c=4D$, for the same bimodal frequency distribution \cite{bns}.

\begin{figure}
\centerline{\hbox{\psfig{figure=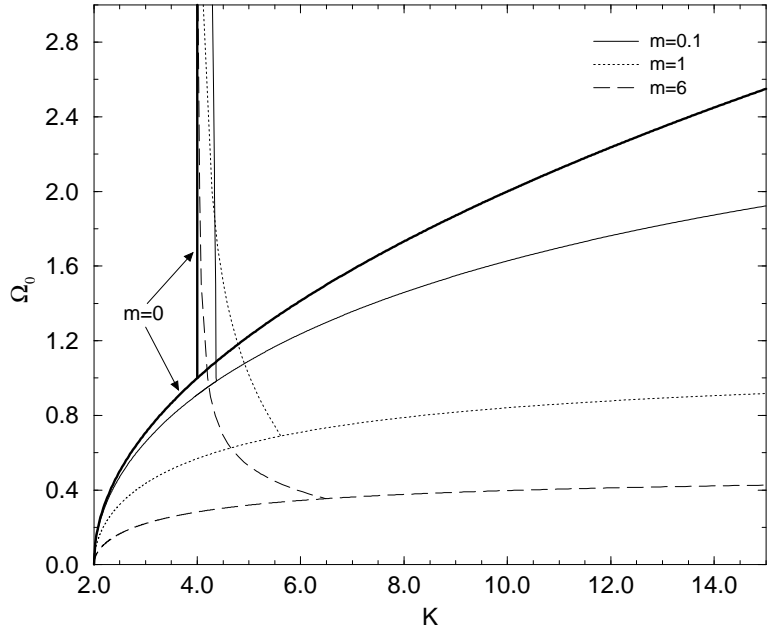,width=12.0cm}}}
\caption{Discrete bimodal frequency distribution:
Stability diagram of incoherence in the
parameter space ($\Omega_0$,
$K$) for different mass values and $D=1$.}
\label{lsb1}
\end{figure}

\begin{figure}
\centerline{\hbox{\psfig{figure=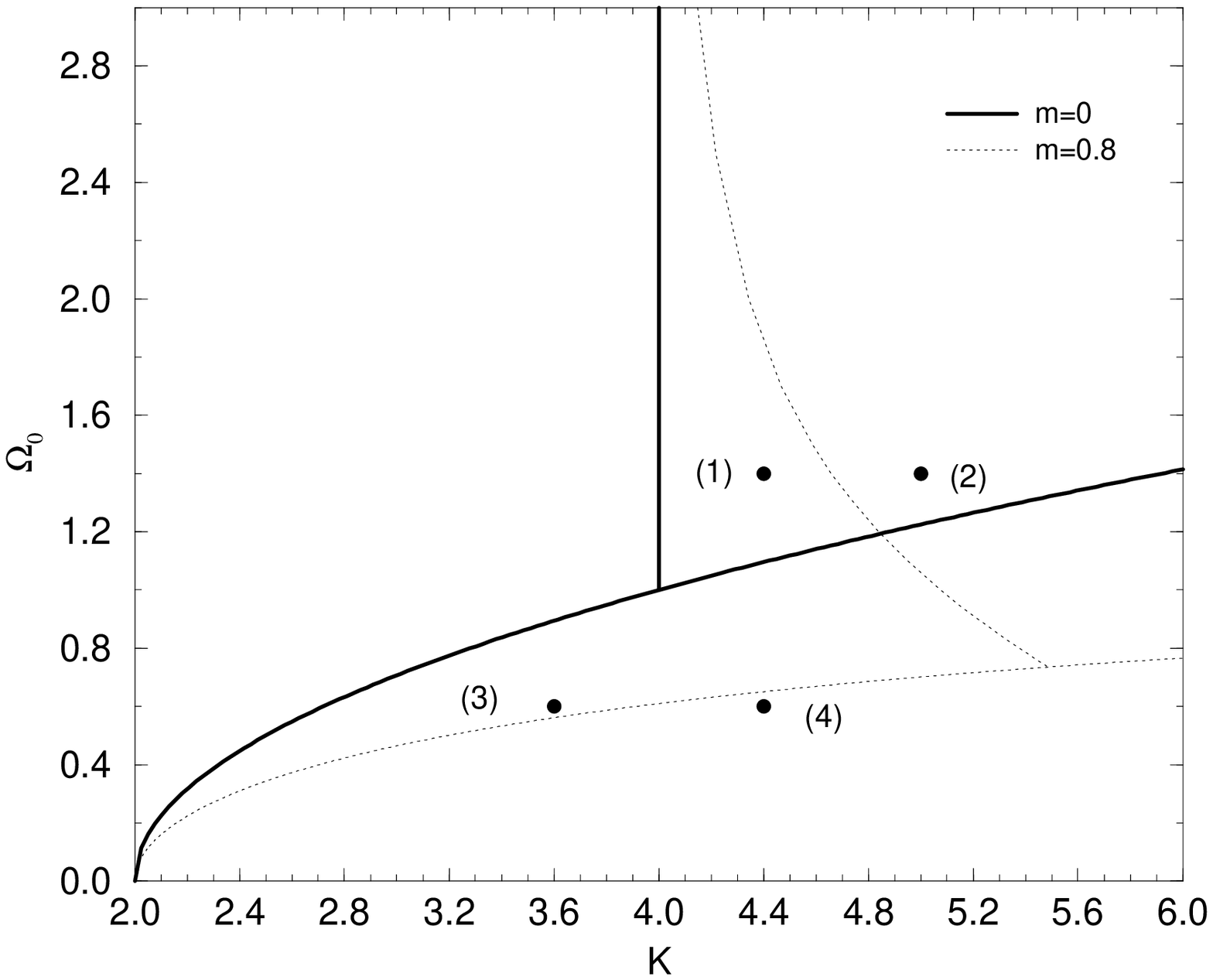,width=11.0cm}}}
\centerline{\hbox{\psfig{figure=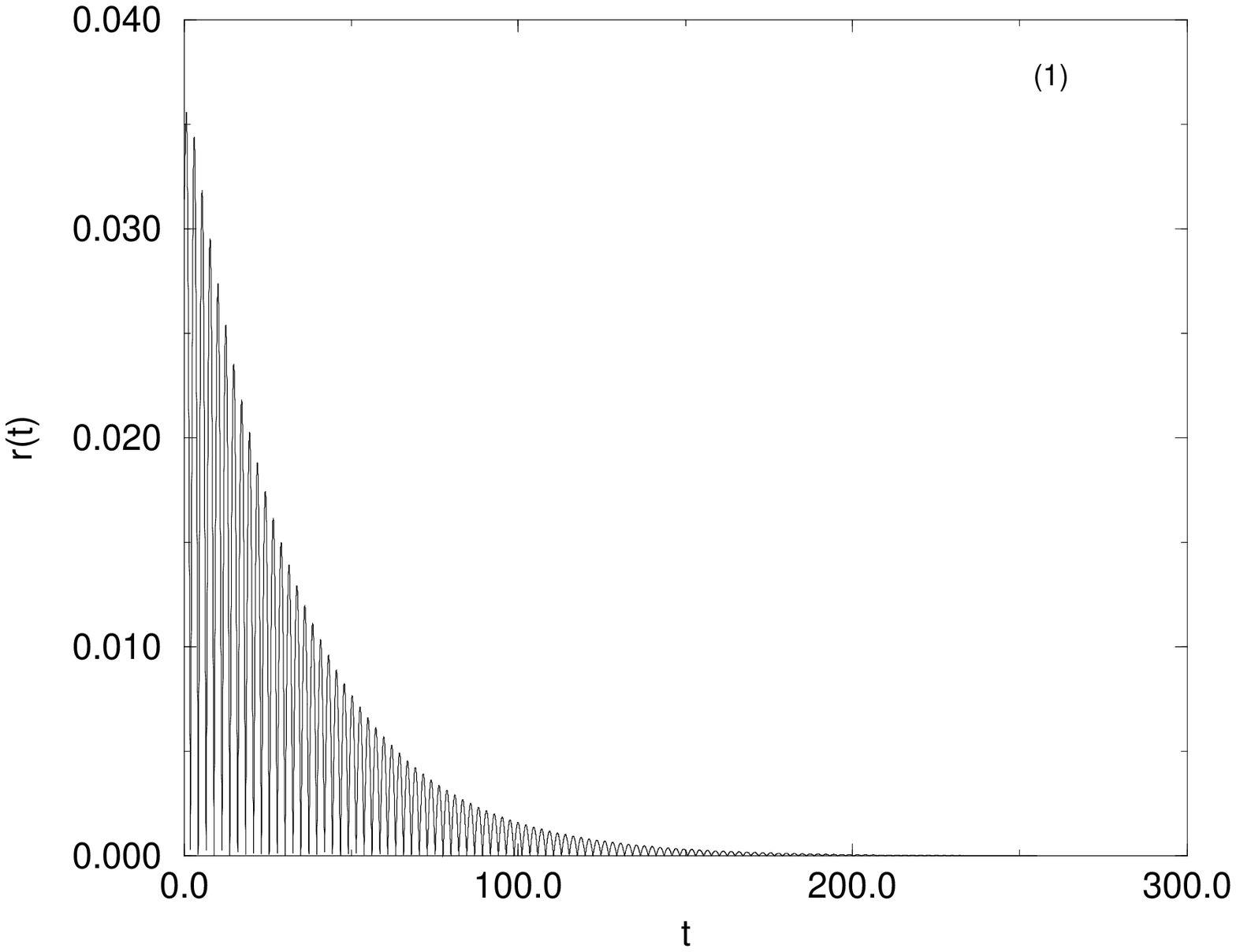,width=7.0cm}}
\hbox{\psfig{figure=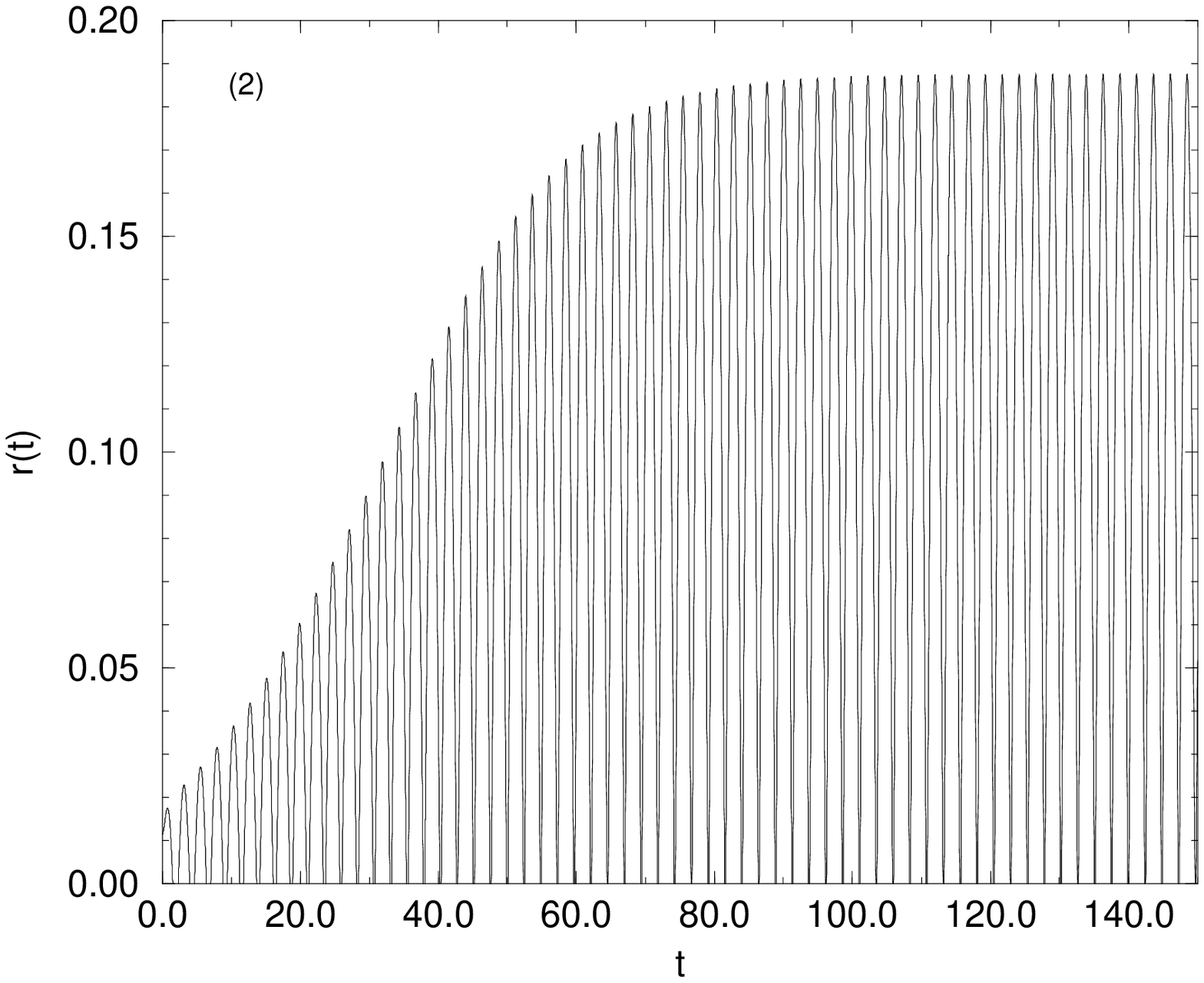,width=7.0cm}}}
\centerline{\hbox{\psfig{figure=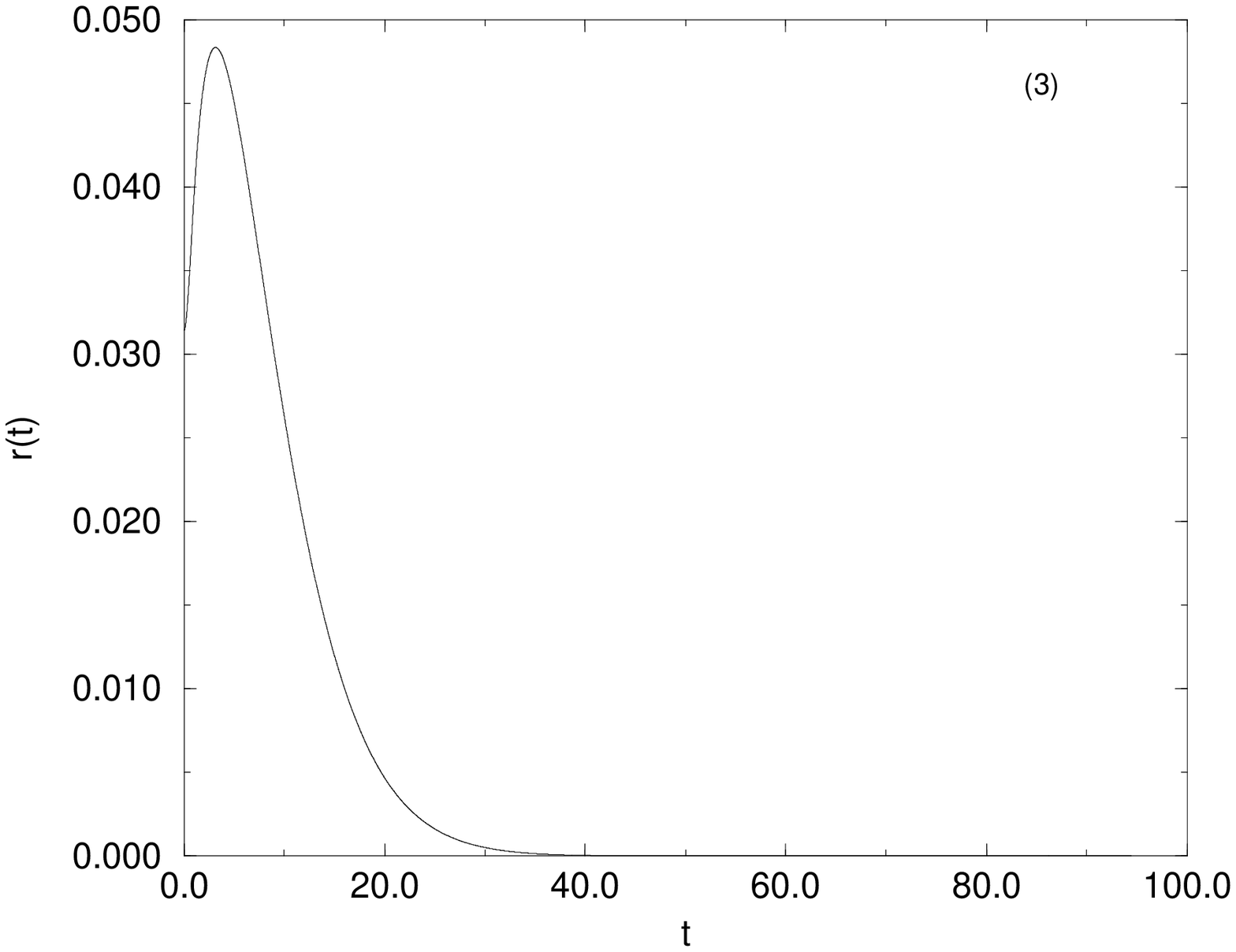,width=7.0cm}}
\hbox{\psfig{figure=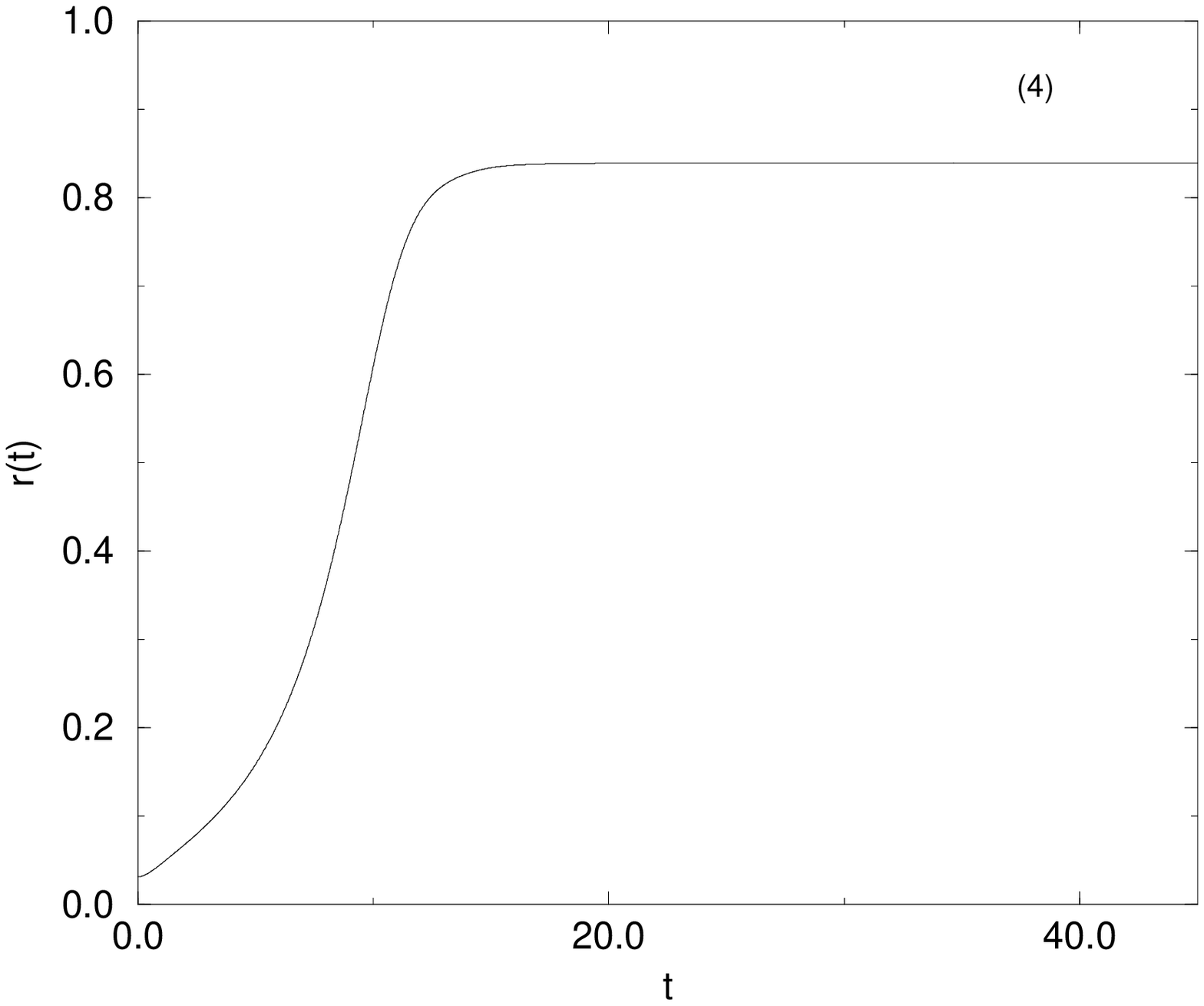,width=7.0cm}}}
\caption{Discrete bimodal frequency distribution:
Stability diagram of incoherence in the parameter
space ($\Omega_0$, $K$) for $m=0.8$ and $D=1$.
The amplitude of the order parameter as a function
of time is displayed at the points marked by: (1)
$K=4.4$, $\Omega_0=1.4$; (2) $K=5$,
$\Omega_0=1.4$; (3) $K=3.6$, $\Omega_0=0.6$; (4)
$K=4.4$, $\Omega_0=0.6$. }
\label{lsb2}
\end{figure}

Figure \ref{lsb1} shows the stability diagram,
which is obtained from Eq.\ (\ref{beta12}) with
Re$\lambda=0$, for $D=1$ and three different mass 
values $m=0.1,\, 1,\, 6$. The stability diagram 
of the Kuramoto model ($m=0$) is also depicted. 
Notice how the curves corresponding 
to differerent masses tend to $K=4D$ as
$\Omega_0\to\infty$, as expected. Fig.\
\ref{lsb2} displays the evolution of the order
parameter amplitude in different regions of the
stability diagram corresponding to $m=0.8$ and
$D=1$.
\bigskip

In conclusion, increasing $m$, $D$, $\varepsilon$
(inertia, noise, and frequency spread) makes it
more difficult to synchronize the oscillator
population via stationary bifurcations from
incoherence.

\section{Mode-coupling equations and stationary
solutions}
Inspired by the previous linear stability
analysis, we shall expand the distribution
function using a basis of parabolic cylinder
functions (or, equivalently, Hermite polynomials)
of unit mean square norm \cite{risken},
\begin{eqnarray}
\rho(\theta,\omega,\Omega,t) = \left(\frac{2\pi
D}{m}\right)^{-{1\over 4}} e^{-{m\omega^{2}\over
4D}} \sum_{n=0}^{\infty} c_n(\theta,\Omega,t)
\psi_{n}(\omega). \label{expansion}
\end{eqnarray}
In (\ref{expansion}), we have defined
\begin{eqnarray}
\psi_n(\omega) = \left(n!\,\sqrt{\frac{2\pi
D}{m}}\right)^{- {1\over 2}}
D_n\left(\sqrt{\frac{m}{D}}\,\omega
\right)\label{s1}\\
= \left(n!\, 2^n \sqrt{\frac{2\pi
D}{m}}\right)^{-1/2}
H_n\left(\sqrt{\frac{m}{2D}}\,\omega\right)\,
e^{-m\omega^2/4D} \nonumber,
\end{eqnarray}
so that $\int_{-\infty}^{\infty}\psi_n\psi_p\,
d\omega=\delta_{np}$. The functions
$c_n(\theta,\Omega,t)$ are $2\pi$-periodic in
$\theta$, and we have
\begin{equation}
\int_0^{2\pi} c_0(\theta,\Omega,t)\, d\theta =1,
\end{equation}
as it follows from the normalization of $\rho(
\theta,\omega,\Omega,t)$.

We shall find a system of mode-coupling equations
for the coefficient functions $c_n$. Then we shall
try to find stationary solutions for different
frequency distributions $g(\Omega)$. This is not
so easy in the general case, so that we shall
follow a standard approach: We shall recognize in
the system of equations for the coefficient
functions a particular stationary solution
corresponding to incoherence. Then we shall try
to find a bifurcation equation for other
stationary solutions branching off from
incoherence. We shall see that even this requires
different approximation schemes in order to
succeed.

\subsection{Mode-coupling equations}
Let us insert Eq.\ (\ref{expansion}) into the Fokker-Planck
equation (\ref{fpe}). We then obtain the following hierarchy
of coupled partial differential equations for
$c_n(\theta,\Omega,t)$,
\begin{eqnarray}
\frac{\partial c_0}{\partial t}=-\sqrt{D\over m}
\frac{\partial
c_1}{\partial\theta},\label{0systemc}\\
\frac{\partial c_1}{\partial t} = -{\cal L} c_0 -
\frac{1}{m} c_1 -\sqrt{2}\sqrt{\frac{D}{m}}
\frac{\partial c_2}{\partial\theta},
\label{1systemc}\\
\vdots \nonumber\\
\frac{\partial c_n}{\partial t}=-\sqrt{n}{\cal L}
c_{n-1} - \frac{n}{m}c_n -\sqrt{n+1}
\sqrt{\frac{D}{m}} \frac{\partial c_{n+1}}
{\partial\theta}.   \label{systemc}
\end{eqnarray}
Here we have defined the operator
\begin{eqnarray}
{\cal L} f = \sqrt{D\over m}\, \left[
\frac{\partial}{\partial\theta}-
\frac{\Omega+K\,r\sin(\psi-\theta)}{D}
\right]\, f.  \label{oper}
\end{eqnarray}
In terms of the functions $c_n$, the equation for
the order parameter becomes
\begin{eqnarray}
r\,e^{i\,\psi}=\int_0^{2\pi}\int_{-\infty}^{+\infty}
e^{i\,\theta}\,c_0(\theta,\Omega,t)g(\Omega) \,d\Omega\,d\theta.
\label{rc0jer}\\
r\, \sin(\psi-\theta) =
\int_0^{2\pi} \int_{-\infty}^{+\infty} 
\sin(\phi-\theta)\, c_0(\phi,\Omega,t)g(\Omega)\,d\Omega\,d\phi\,.
\end{eqnarray}

\subsection{Incoherence and bifurcating
stationary solutions}
The {\it incoherent} $\theta$-independent
solution, $c_n=C_n(\Omega)$, depends only on
$\Omega$. It can readily be obtained from the
above hierarchy of equations, by ignoring all
derivatives. The result is
\begin{equation}
C_n(\Omega)=\frac{1}{2\pi}\frac{1}{\sqrt{n!}}
\left({\frac{m}{D}}\right)^{n/2}\,\Omega^n.
\label{cnbar}
\end{equation}
Inserting this into the expansion in Eq.\
(\ref{expansion}), we indeed recover Eq.\
(\ref{incoherent}). In particular setting
$\Omega=0$, this yields the simplest incoherent
solution, $C_n=1/(2\pi)$, corresponding to the
unimodal frequency distribution. Other interesting
stationary solutions are partially synchronized
distributions, which depend on $\theta$. Notice
that eq. (\ref{0systemc}) implies that $c_1$ does
not depend on $\theta$.

In the following we analyze stationary solutions
bifurcating from incoherence for different
frequency distributions.

\bigskip
\noindent
(a) {\it Discrete unimodal frequency distribution},
$g(\Omega)=\delta(\Omega)$.

Let us look for {\it stationary} solutions with
finitely many nonzero coefficients $c_n = 0$ for
$n>N$ and $g(\Omega) =\delta(\Omega)$. Eq.\
(\ref{systemc}) yields ${\cal L}c_N =0$, and
therefore, $c_N = K_N\, e^{\frac{K}{D} \,r
\cos(\psi-\theta)}$ ($K_N=$constant). Inserting
this result in (\ref{systemc}) for $n=N-1$, we
find ${\cal L}c_{N-1} = -\sqrt{N}c_N/m$. The
corresponding solution is not periodic in
$\theta$ unless $K_N=0$. Repeating this argument, we obtain the
stationary solution
\begin{equation}
c_0(\theta)=\frac{e^{\frac{K}{D} \,r
\cos(\psi-\theta)}} {\int^{2\pi}_{0} e^{\frac{K}{D} \,r
\cos(\psi-\theta)}d\theta},
\end{equation}
where $c_n=0$ for $n>0$ and the normalization
condition has been used. Observe that the
resulting distribution, $\rho(\theta,\omega)=
\left(\frac{2\pi\,D}{m}\right)^{1/4} c_0(\theta)
\psi_{0}(\omega)\,e^{-m\omega^2/4D}$, is factorized with respect to
its two  arguments, $\theta$ and $\omega$, cf.\ \cite{acebron1}. It is
remarkable that
$c_0(\theta)$ is independent of $m$, and
coincides with that obtained for  the Kuramoto
model. Therefore, from (\ref{rc0jer}) the order
parameter does not depend on inertia, and the
bifurcation diagram for $r$ is exactly the same as
that for the Kuramoto model.

\bigskip
(b) {\it General frequency distributions.}

For general frequency distributions, we could try
to find stationary solutions of the mode-coupling
equations (\ref{systemc}) which bifurcate from
incoherence. It is however more direct to work
with the stationary Fokker-Planck equation
(\ref{fpe}) as follows. We consider stationary
solutions as functions of a fixed value of the
synchronization parameter $r$. Then we expand
these solutions in power series of $r$ and write
a hierarchy of equations for the coefficient
functions. The first coefficient function should
be the incoherent solution of synchronization
parameter $r=0$. Inserting the power series
probability density function into Eq.\
(\ref{order}), we find the amplitude equation for
stationary solutions bifurcating from incoherence.
This procedure is explained in Appendix B. We
quote here the result:
\begin{equation}
r=\frac{K\,r}{2D}\alpha +\frac{(K\,r)^3}{6}\beta
+ O((K\,r)^4),
\label{ralpha}
\end{equation}
where
\begin{eqnarray}
\alpha=e^{m\,D}
\left[\int_{-\infty}^{\infty}
\frac{1}{1+\frac{\Omega^2}{D^2}}g(\Omega)\,d\Omega \right.\nonumber\\
\left.+\sum_{p=1}^{\infty}
 \frac{(-m\,D)^{p}(1+\frac{p}{m\,D})^{2}}{p!}
\right.\nonumber\\
\left.\times \int_{-\infty}^{\infty}\frac{1}{(1+\frac{p}{m\,D})^2+
\frac{\Omega^2}{D^2}}g(\Omega)\,d\Omega
\right],
\label{alphateo}
\end{eqnarray}
and $\beta$ can be calculated numerically by
solving a system of differential equations. As
the inertia vanishes, $mD\to 0+$, $\alpha$
becomes
\begin{eqnarray}
\alpha=\int_{-\infty}^{\infty} \frac{g(\Omega)\,
d\Omega}{1+ \frac{\Omega^2}{D^2}} -\frac{m}{D}
\int_{-\infty}^{\infty} \frac{\Omega^2}{1+
\frac{\Omega^2}{D^2}} g(\Omega) d\Omega +
O(m^2 D^2).
\label{alphabeta1}
\end{eqnarray}
Notice that $\alpha$ in Equation (\ref{alphateo})
coincides with $2D/K$ in Eq.\ (\ref{dispersion3})
provided $\lambda=0$ and $g(\Omega)=g(-\Omega)$.
This means that the critical coupling for
bifurcation towards stationary synchronized states
is obtained at $K\alpha=2D$, no matter what the
symmetric frequency distribution may be.

\begin{figure}
\centerline{\hbox{\psfig{figure=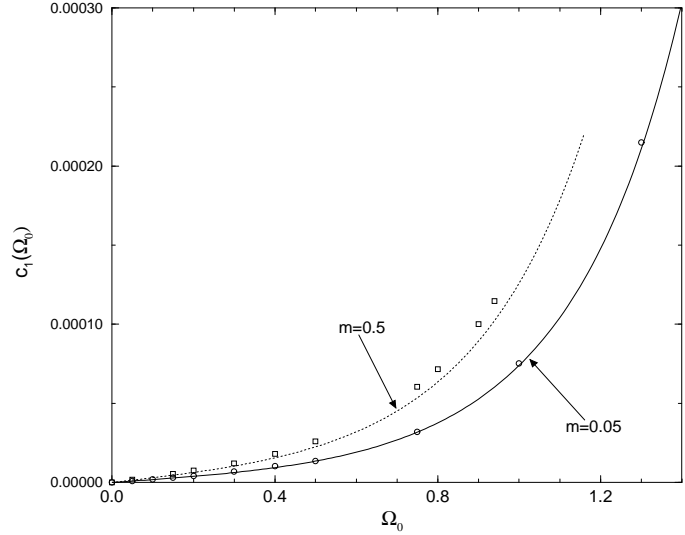,width=9.0cm}}}
\caption{Coefficient $c_1(\Omega_0)$ corresponding to the
approximate mode-coupling stationary solution for the
discrete bimodal frequency distribution and two different
values of $m$. The solid ($m=0.05$) and dotted ($m=0.5$) lines
correspond to the 3-mode approximation, whereas squares and
circles are obtained from direct numerical simulation.
Notice that $c_1$ is even: $c_1(-\Omega_0)=c_1(\Omega_0)$.}
\label{ss0}
\end{figure}

Before interpreting the amplitude equation
(\ref{ralpha}), we shall outline a procedure to
obtain its coefficients based upon an uncontrolled
closure assumption which is accurate for small
values of $mD$. Consider the expansion
(\ref{expansion}). We expect that the coefficients
$c_n$ of stationary solutions close to incoherence
do not differ much from the coefficients of the
latter. In view of the functional form
(\ref{cnbar}), we anticipate that the
coefficients $c_n$ approach zero as $n\to \infty$
{\em faster for smaller values of the mass}. Thus
we shall now consider stationary solutions for
general frequency distributions, such that
$c_n=0$ in (\ref{systemc}), for all $n\ge 3$. By
solving equation (\ref{systemc}) for such a
stationary solution with $n=2$, we obtain
\begin{equation}
c_2(\theta,\Omega)=\sqrt{\frac{m}{2\,D}}
[\Omega+K\,r\, \sin(\psi-\theta)]\, c_1(\Omega).
\label{c2}
\end{equation}
We now insert this expression into
(\ref{1systemc}). The resulting equation is
solved for a $c_0$, which is $2\pi$-periodic in
$\theta$ and obeys the normalization condition
(\ref{norma}). We find
\begin{eqnarray}
c_0(\theta,\Omega) = {e^{\frac{K\,r}{D} \cos
(\psi-\theta)}
\varphi(\theta,\Omega)\over Z(\Omega)}
\label{c0stat} \\
\varphi(\theta,\Omega) = \int_0^{2\pi} [1 - m K r
\cos (\psi-\theta-\eta)]\nonumber\\
\times e^{-\frac{1}{D}[\Omega \eta+ K\,r\,\cos
(\psi-\theta-\eta)]} d\eta, \label{varphi}\\
Z(\Omega) = \int_0^{2\pi} e^{\frac{K\,r}{D}\cos
(\psi-\theta)} \varphi(\theta,\Omega)\, d\theta,
\label{zeta}\\
\end{eqnarray}
In (\ref{c0stat}), $r$ should be determined so
that
\begin{eqnarray}
r=\int_0^{2\pi}\int_{-\infty}^{\infty}  c_0(\theta,\Omega)  \cos
(\psi-\theta)g(\Omega)\,d\Omega
d\theta,
\label{rc0}
\end{eqnarray}
holds.  The function $c_1(\Omega)$ can be obtained
by integrating (\ref{1systemc}) with respect to 
$\theta$ and using the normalization condition
for $c_0$ together with the $2\pi$-periodicity in
$\theta$ of $c_0$ and $c_2$: 
\begin{equation}
c_1=\frac{1}{2\pi}\sqrt{\frac{m}{D}}\left[
\Omega+K\,r\,\int_0^{2\pi}
\sin(\psi-\theta)c_0(\theta,\Omega)\,d\theta
\right].\label{c1a}
\end{equation}
Then, the stationary distribution can be
approximated by
\begin{eqnarray}
\rho(\theta,\omega,\Omega) \approx
\left(\frac{2\pi D}{m}\right)^{1/4}
[c_0(\theta,\Omega) \psi_0(\omega) +
c_1(\Omega)\psi_1(\omega) \nonumber\\
+ c_2(\theta,\Omega)\psi_2(\omega)]\,
e^{-m\omega^2/4D}.
\label{sdis}
\end{eqnarray}
Notice that a nonvanishing $c_1(\Omega)$ in
(\ref{expansion}) implies that the probability
density is no longer even in $\omega$ ($\psi_1$
is an odd function of $\omega$). By (\ref{c1a}),
this occurs in the synchronized phase for the
case of nonidentical oscillators. For the bimodal
frequency distribution, Fig.\ \ref{ss0} shows
$c_1$ as function of $\Omega_0$ for two different
values of $m$. Both the approximate expression
(\ref{c1a}) and results of direct numerical
simulations are depicted. Notice that the
agreement between our approximation and the
numerical result improves as $m$ decreases. Thus
we observe that for each fixed nonzero
$\Omega$ and each fixed $\theta$, the distribution
function is no longer peaked at $\omega=0$.
However, the {\em instantaneous frequency
distribution}, defined by 
$$\int_0^{2\pi} \int_{-\infty}^{\infty}
\rho(\theta,\omega,\Omega)\,  g(\Omega)
d\Omega\,d\theta\,,$$
may turn out to be even in $\omega$. This is
certainly true for the approximate stationary
distribution (\ref{sdis}), for 
$$
\int_{-\infty}^{\infty} c_1(\Omega) g(\Omega) 
d\Omega = \sqrt{{m\over D}}\, {Kr\over
2\pi}\, \int_0^{2\pi}\int_{-\infty}^{\infty}
c_0(\theta,\Omega) \sin(\psi-\theta)\, g(\Omega) 
d\Omega d\theta = 0,
$$
as it follows from the definition of the order 
parameter and the expansion (\ref{expansion}). 

Another indication that the exact instantaneous
frequency distribution may be even in $\omega$ is
that the average frequency tends to zero as $t\to
+\infty$. This would occur if the stable
stationary distribution is even in $\omega$
(although, admittedly, distributions which are
not even in $\omega$ may have zero mean). The
result can be shown directly from the
Fokker-Planck equation (\ref{fpe}). We multiply
that equation by $\omega\, g(\Omega)$ and
integrate with respect to all variables. Then we
obtain the following equation for the mean value 
$\langle \omega\rangle$:
$$ {d\over dt}\langle \omega\rangle + {\langle
\omega\rangle\over m} = \int_{-\infty}^{\infty}
\Omega g(\Omega) d\Omega.$$ 
The right hand side of this expression is zero if
$g(-\Omega) = g(\Omega)$. Then $\langle
\omega\rangle$ tends to zero exponentially fast
as $t\to + \infty$. 

Let us now go back to the problem of obtaining the
coefficients in the amplitude equation
(\ref{ralpha}). Let us expand (\ref{c0stat}) in
powers of $r$ and insert the result
in (\ref{rc0}). We obtain an approximation to the
coefficients of the amplitude equation
(\ref{rc0}). $\alpha$ is again given by
(\ref{ralpha}), and the expression for $\beta$ is
\begin{eqnarray}
\beta = \int_{-\infty}^{\infty}
\left[ \frac{3\frac{\Omega^2}{D^2} -\frac{3}{2}
+\frac{39}{4} m\frac{\Omega^2}{D}}
{{(1+\frac{\Omega^2}{D^2})}^{2}(4+\frac{\Omega^2}{D^2}) D^{3}}
\right. \nonumber\\
\left. + \frac{3 m^{2} \Omega^{2}
\left(1+\frac{\Omega^2}{2D^2}\right)}{2 D^{3}
\left(1+ \frac{\Omega^2}{D^2}\right)^2}\right]\,
g(\Omega) d\Omega .
\label{alphabeta2}
\end{eqnarray}

We now interpret the amplitude equation
(\ref{ralpha}) in the usual way. Notice that the
nonzero solution is approximately given by
$$Kr\sim \sqrt{{3(2D-K\alpha)\over K\beta D}}\,.
$$
Then the critical value of $K$ is $K^{*}=
2D/\alpha$, and the sign of $\beta$ determines
the direction of the bifurcating branch of
stationary solutions. Assume $\alpha>0$. Then the
bifurcating stationary solution exists for $K>
K^{*}$ if $\beta<0$ ({\it supercritical
bifurcation}). If $\beta>0$, we have a {\it
subcritical bifurcation} and the partially
synchronized stationary solution exists for $K<
K^{*}$. The critical coupling tends
to infinity as $\alpha\to 0+$, and the
bifurcation does not occur at positive couplings
if $\alpha<0$. Given the formulas (\ref{alphateo})
or (\ref{alphabeta1}), $\alpha$ could become
negative, depending on the value of $m$. Let us
now analyze the bifurcation diagram corresponding
to two different frequency distributions.

\begin{figure}
\centerline{\hbox{\psfig{figure=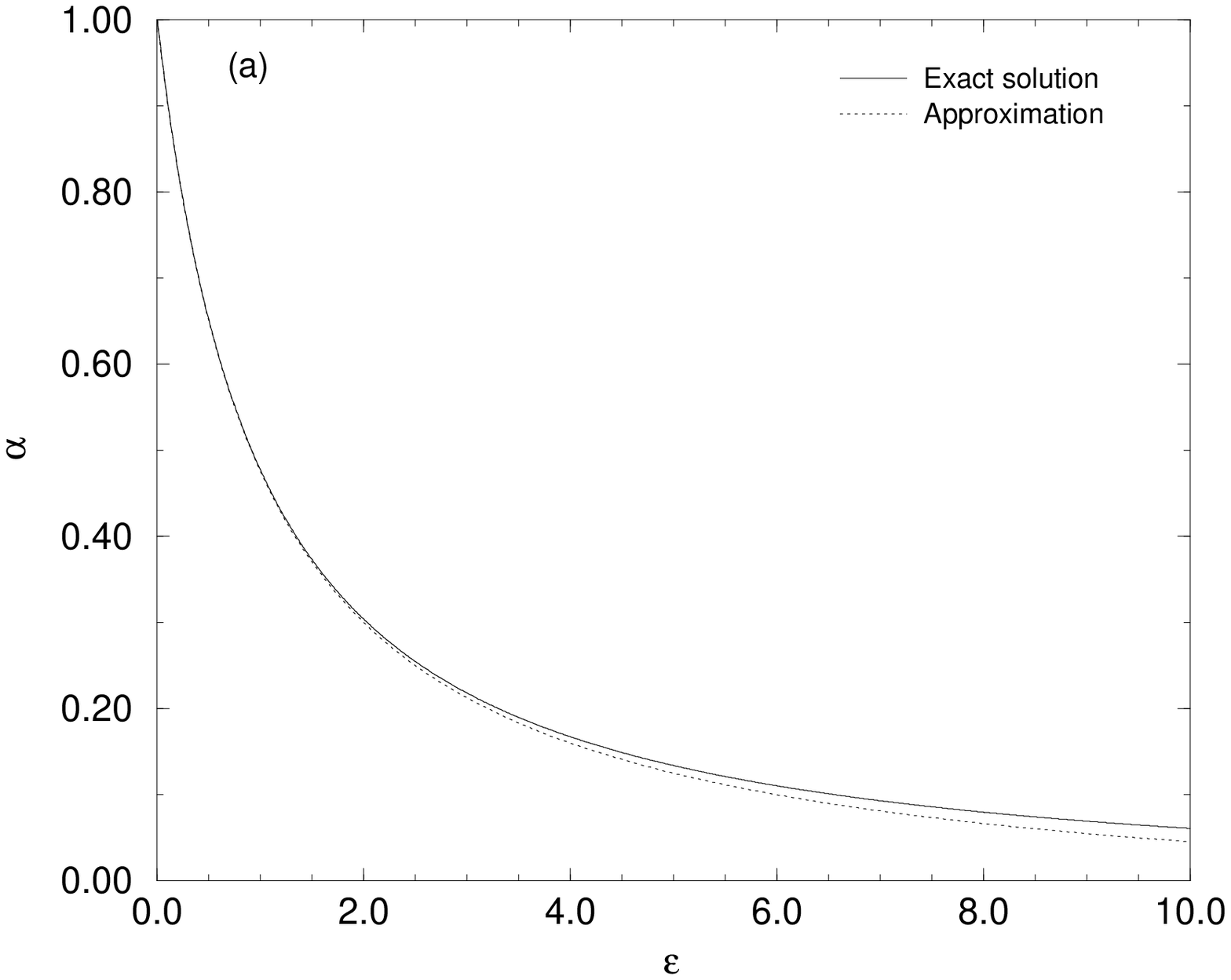,width=8.0cm}}
\hbox{\psfig{figure=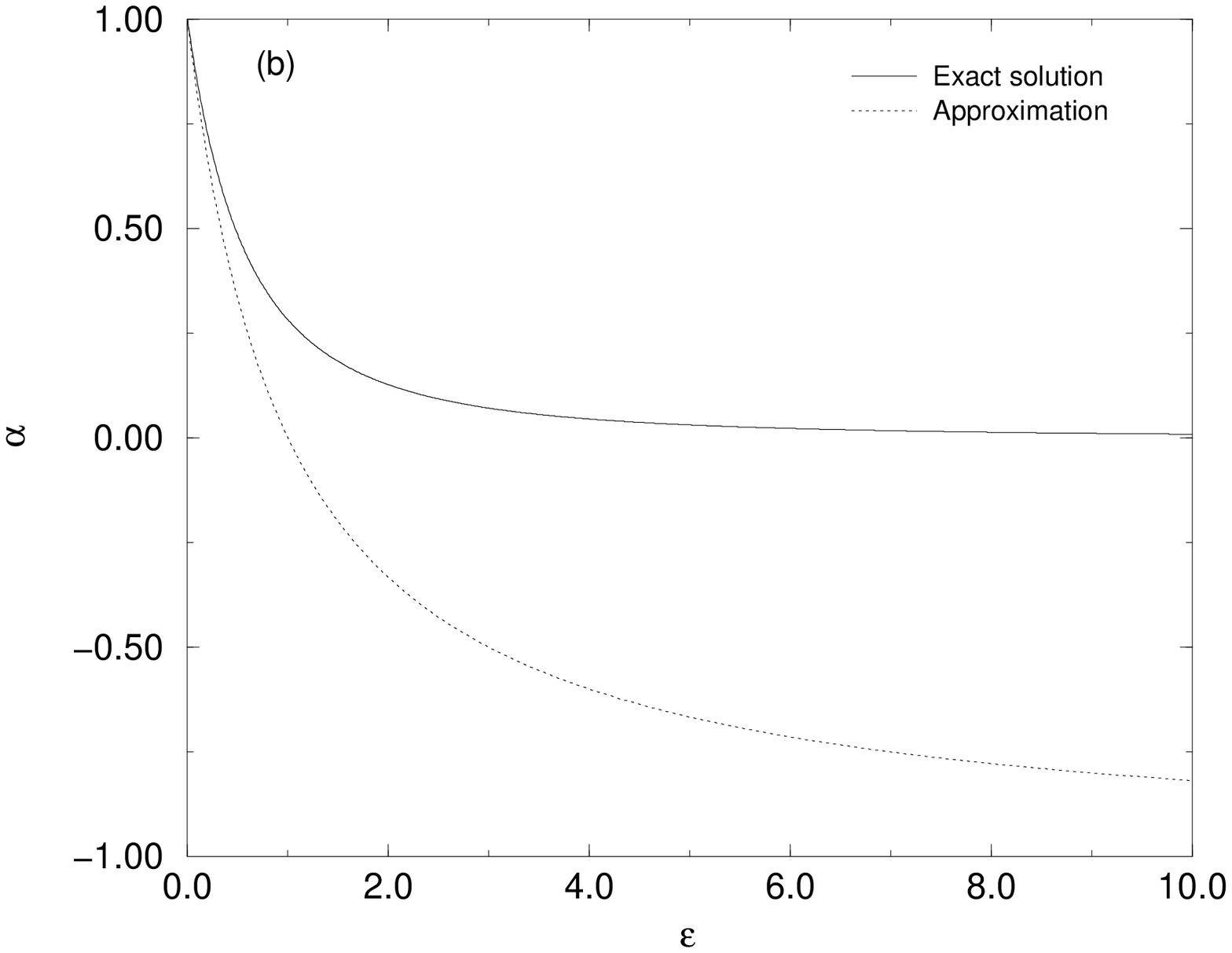,width=8.0cm}}}
\caption{Coefficient $\alpha$ as a function of the frequency
spread $\varepsilon$ for the stationary solution in the case
of unimodal Lorentzian frequency distribution. The
exact result and that of the 3-mode approximation
are compared. Other parameter values are $D=1$, and (a)
$m=0.05$; (b) $m=1$. Note that the approximate result 
improves as $m$ and $\varepsilon$ decrease.}
\label{sslor1}
\end{figure}

\begin{figure}
\centerline{\hbox{\psfig{figure=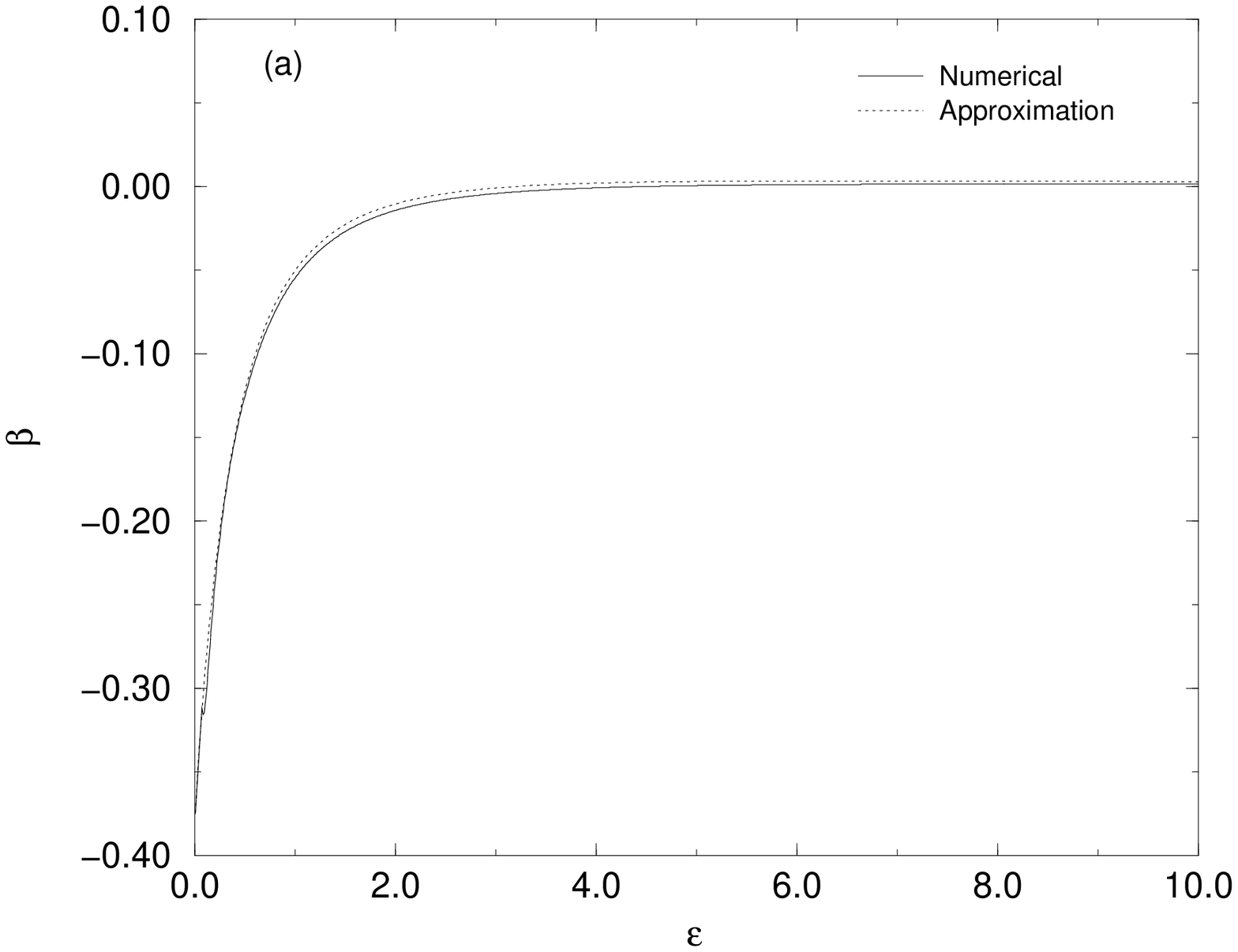,width=8.0cm}}
\hbox{\psfig{figure=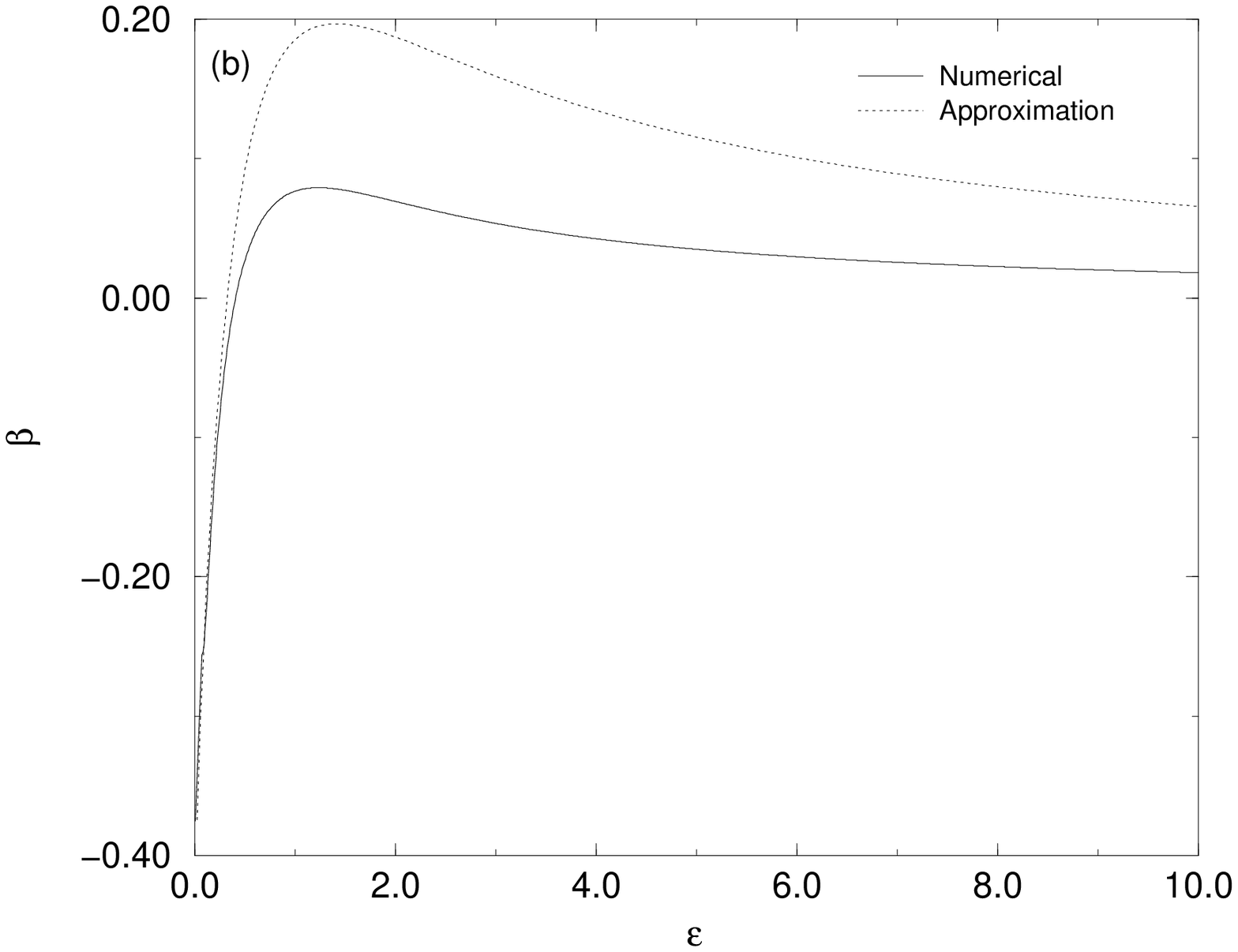,width=8.0cm}}}
\caption{Coefficient $\beta$ as a function of the frequency
spread $\varepsilon$ for the stationary solution in the case
of unimodal Lorentzian frequency distribution. The
exact result and that of the 3-mode approximation
are compared. Other parameter values are $D=1$, and (a)
$m=0.05$; (b) $m=1$. Note that the approximate result 
improves as $m$ and $\varepsilon$ decrease.}
\label{sslor2}
\end{figure}

\begin{figure}
\centerline{\hbox{\psfig{figure=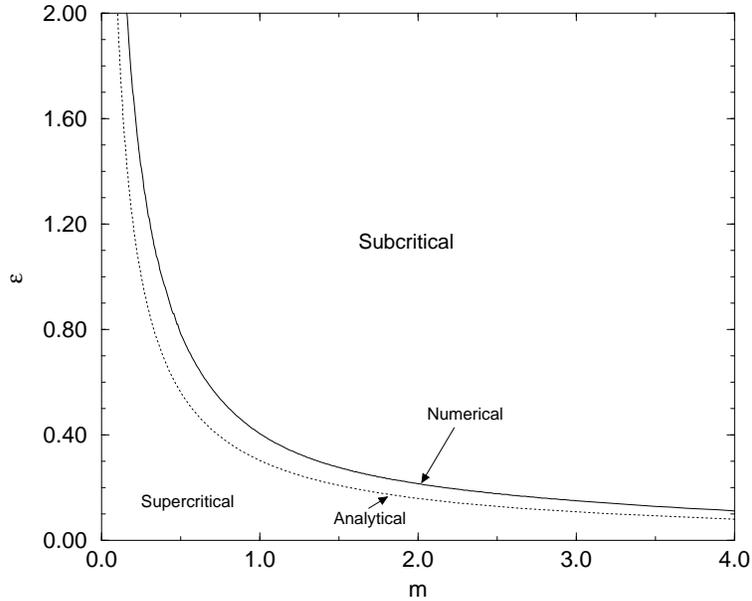,width=10.0cm}}}
\caption{Stationary solution, Lorentzian frequency 
distribution: Relation between frequency spread and mass at
the  critical line $\beta=0$. This line separates the regions
in parameter space on which the synchronization transition
from incoherence is either a sub or a supercritical
bifurcation. The numerically evaluated curve is
compared to the approximate result from 3-mode truncation.}
\label{sslor3}
\end{figure}

\begin{figure}
\centerline{\hbox{\psfig{figure=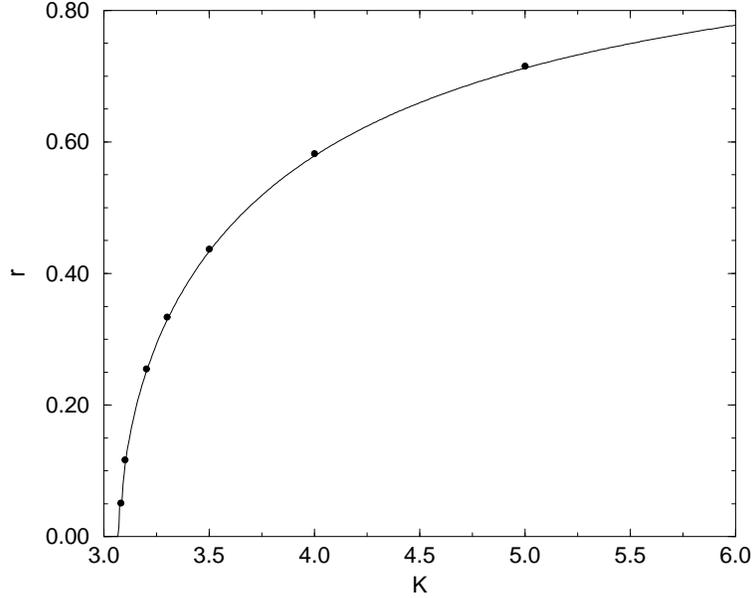,width=10.0cm}}}
\caption{Bifurcation diagram for a supercritical bifurcation
from incoherence to a synchronized stationary solution,
corresponding to the unimodal Lorentzian frequency
distribution and parameter values of $m=0.05$, $D=1$,
$\varepsilon=0.5$. The amplitude of the order parameter is
represented as function of $K$. The continuous line is 
the analytical approximation and the dots are obtained by
direct numerical simulation. }
\label{sslor4}
\end{figure}

\begin{figure}
\centerline{\hbox{\psfig{figure=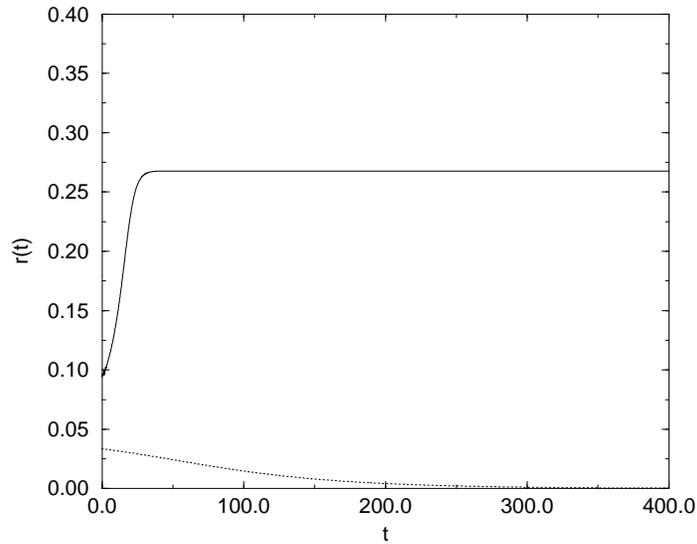,width=10.0cm}}}
\caption{Evolution of the amplitude of the order parameter
for two different initial conditions, unimodal Lorentzian
frequency distribution and parameter values $m=0.05$,
$D=1$, $\varepsilon=5$, and $K=14.6$ (in the region of
subcritical bifurcation from incoherence).  The simulations
illustrate the existence of bistability between incoherence
and synchronized stationary solution. }
\label{sslor5}
\end{figure}

\begin{figure}
\centerline{\hbox{\psfig{figure=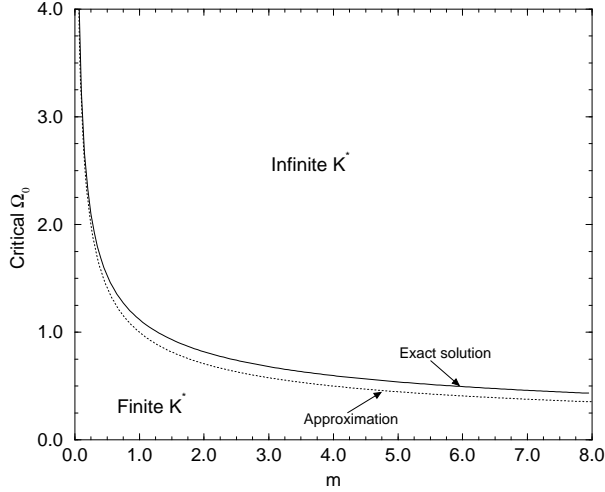,width=8.0cm}}}
\caption{Stationary solution, bimodal case: 
Critical frequency $\Omega_0^{\infty}$, at which
$\alpha=0$, as a function of mass. Below the
critical line, stationary synchronized states
bifurcate from incoherence at a finite $K^{*}$.
Above  the critical line, $K^{*}=\infty$, at
which value the branch of synchronized states
bifurcates subcritically from incoherence. The
exact result for the critical line and that
obtained from 3-mode truncation are also
compared.}
\label{ssb1}
\end{figure}

\begin{figure}
\centerline{\hbox{\psfig{figure=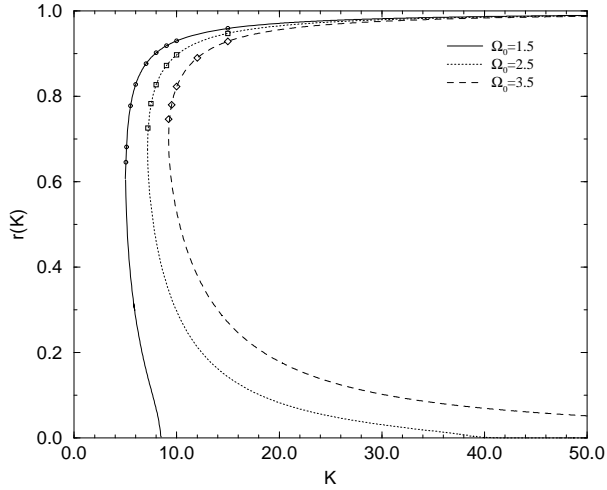,width=8.0cm}}}
\caption{Stationary solution, bimodal case: 
Amplitude of the order parameter, $r$, in the
stationary synchronized  state as a function of
the coupling strength, $K$. We have represented
results from the 3-mode truncation (lines) and
from numerical simulations (dots, squares,
circles) for three differents values of
$\Omega_0$. Other parameter values are $m=0.1$
and $D=1$. Notice that $K^{*}$ increases with
$\Omega_0$, and it becomes infinity for $\Omega_0
=3.5$, larger than the critical value
$\Omega_0^{\infty}\approx 3.16$, as expected.
This figure illustrates that synchronization
issues forth from incoherence as a subcritical
bifurcation at
$K^{*}=\infty$. }
\label{ssb2}
\end{figure}

\begin{figure}
\centerline{\hbox{\psfig{figure=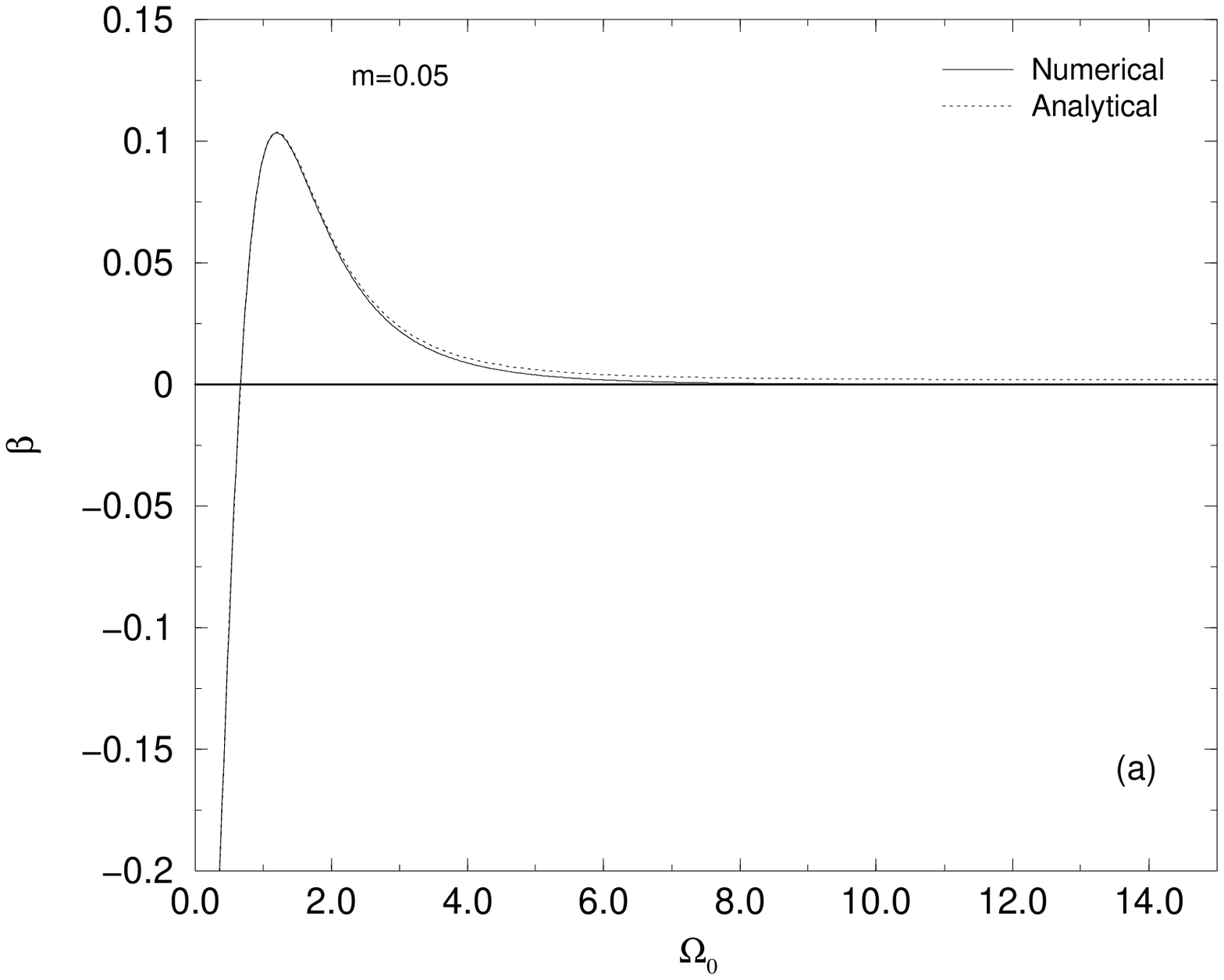,width=7.0cm}}
\hbox{\psfig{figure=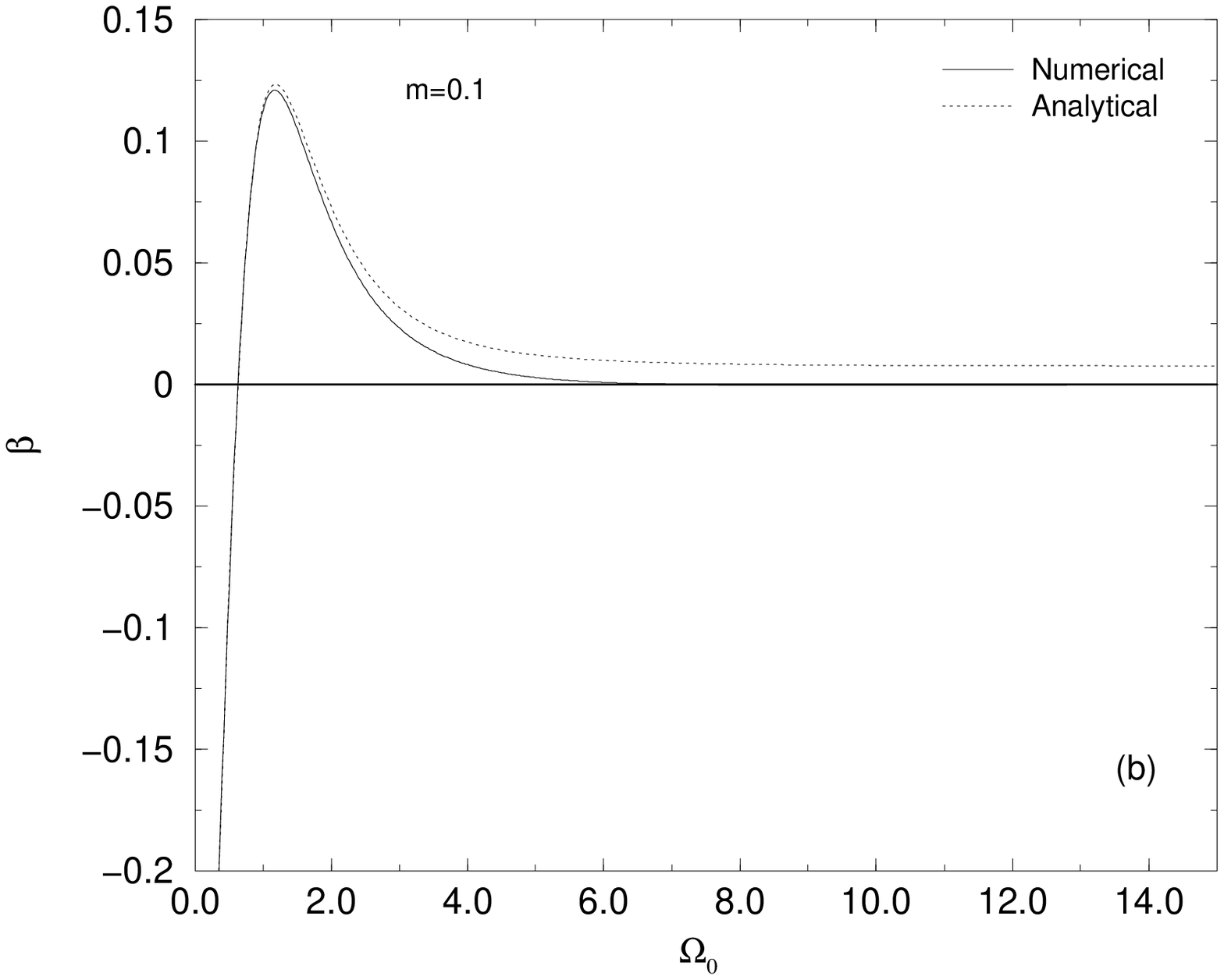,width=7.0cm}}}
\caption{Stationary solution, bimodal case: 
Coefficient $\beta$ as a function of $\Omega_0$.
We have compared direct numerical evaluation of
$\beta$ (obtained by solving numerically the
system of ordinary differential equations in
Appendix B for the case n=3), and Equation 
(\ref{alphabeta2}) for the discrete bimodal
frequency distribution. Other parameter values
are $D=1$ and: (a) $m=0.05$, and (b) $m=0.1$.}
\label{ssb3}
\end{figure}

\begin{figure}
\centerline{\hbox{\psfig{figure=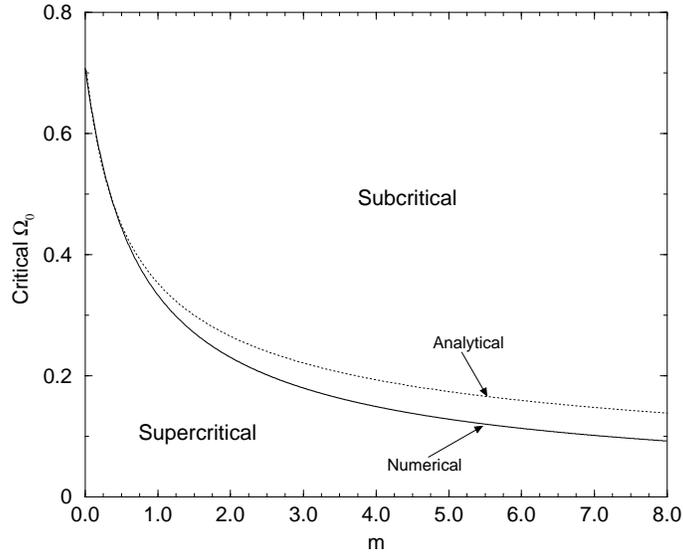,width=9.0cm}}}
\caption{Stationary solution, bimodal case: 
Critical line at which $\beta=0$ in the
$(\Omega_0,m)$ plane. We have displayed both the
direct numerical result and Equation
(\ref{alphabeta3}).} 
\label{ssb4}
\end{figure}

\begin{itemize}
\item{\it Unimodal Lorentzian frequency
distribution}.

For an unimodal Lorentzian frequency distribution,
the coefficients $\alpha$ and $\beta$ of
(\ref{alphabeta1}) and (\ref{alphabeta2}) are
approximately given by
\begin{eqnarray}
\alpha=\frac{D}{D+\gamma}(1-m\,\varepsilon),\\
\beta=\frac{3}{4}\frac{1}{(D+\varepsilon)^2
(2D+\varepsilon)}[-1+\frac{m}{D} \varepsilon(3D+\varepsilon)].\label{lorss}
\end{eqnarray}
Apparently, the sign of $\alpha$ could again be
negative provided $m\varepsilon$ is sufficiently
large. However, evaluation of the exact
expression (\ref{alphateo}) shows that $\alpha> 0$. 
Figure \ref{sslor1} shows $\alpha$ as a 
function of $\varepsilon$ for two different
masses. Notice that the approximate expression
for $\alpha$ fits better the numerical result as
$m$ and $\varepsilon$ decrease. 

In contrast with $\alpha$, $\beta$ can
really change sign for a Lorentzian frequency
distribution. Fig.\ \ref{sslor2} shows the good
qualitative agreement between the
approximate expression for $\beta$ and its
numerical evaluation from the exact equations 
(\ref{z3sys}). 

There is a critical mass $m_c$ ($\varepsilon$ and
$D$ are kept fixed) for which $\beta=0$. This
mass separates the supercritical and subcritical
bifurcation regions. Setting $\beta=0$ in eq.
(\ref{lorss}) yields
\begin{equation}
m_c=\frac{D}{\varepsilon(3D+\varepsilon)}.
\label{criticallor}
\end{equation}
Figure \ref{sslor3} shows the qualitative
agreement between the approximate expression
(\ref{criticallor}), and the numerical result
obtained from the solution of (\ref{z3sys}).

Note that in the massless case, the stationary
solution always branches off supercritically,
independently of $\varepsilon$. In the presence
of inertia and for appropriate values of
$\varepsilon$, we have found a subcritical
bifurcation. This prediction is illustrated by  
Figures \ref{sslor4} (parameters corresponding 
to a supercritical bifurcation) and \ref{sslor5}
(parameters corresponding  to a subcritical
bifurcation). Figure \ref{sslor4} is a typical 
supercritical bifurcation diagram (order
parameter amplitude versus $K$). In the
subcritical case, Figure \ref{sslor5} shows the
different evolution of the order
parameter amplitude for two different
initial conditions. Notice the bistability between
incoherence and the stable synchronized
solution typical of a subcritical bifurcation.

\item {\it Bimodal frequency distribution}

For the bimodal frequency distribution, $\alpha$
may be zero. This means that  there is a critical
frequency $\Omega_{0}^{c}(m)$ above which $K^{*}
= \infty$. In fact, we have shown in  Fig. \ref{lsb1},
looking to the branch corresponding to
$\lambda=0$, that in case of a  bimodal frequency
distribution, it is possible to find some values
of $\Omega_0$ without its corresponding
$K$ in the stability diagram of incoherence, due to the existence of an
horizontal asymptote in the space parameter $(K,\Omega_0)$. This means that
there is not a finite coupling $K^{*}$ where the stationary solution
branches off. On the other hand, this never
occurs
in the bimodal Kuramoto model [$m=0$,
 $K_c/D=2(1+\Omega_0^2/D^2)$], because in this
case there is not such an horizontal asymptote.

Eq.\ (\ref{alphabeta1}) shows that the critical
frequency at which $\alpha=0$ is
\begin{equation}
\frac{\Omega_0^{\infty}}{D}=\sqrt{\frac{1}{m\,D}}.
\end{equation}
Figure \ref{ssb1} compares the previous
approximate expression to the exact critical
frequency obtained from (\ref{alphateo}). Note
that the approximation improves as $m$
decreases. Fig. \ref{ssb2} shows that 
$K^{*}$ increases as $\Omega_0$ does. $K^{*}$
becomes infinity for $\Omega_0\geq
\Omega_0^{\infty}$. 

For $\alpha>0$, there is another important
critical frequency. In this case, the sign of
$\beta$ decides whether the bifurcation is
subcritical ($\beta>0$) or supercritical
($\beta<0$). The sign of $\beta$ depends on $m$
and $\Omega_0$. For small mass, we can use the
approximations (\ref{alphabeta1}) and
(\ref{alphabeta2}) (ignoring terms which are
quadratic in the mass). Then we find that the
critical frequency at which
$\beta=0$ is
\begin{equation}
\frac{\Omega_0^{c}}{D}=\sqrt{\frac{1}{2+
\frac{13}{8}m\,D}}.\label{alphabeta3}
\end{equation}
We have solved numerically the system of
equations (\ref{z3sys}), and compared with the
approximation (\ref{alphabeta2}) for $\beta$. In
Figure \ref{ssb3} the coefficient $\beta$ is
plotted as a function of $\Omega_0$ for different
masses. Similarly, Figure \ref{ssb4} shows how
the critical frequency $\Omega_0^{c}$ varies as a
function of $m$. Note that the analytical
approximation improves as $m$ goes to zero (as
expected). 

Notice that $\Omega_0^{c}<\Omega_0^{\infty}$.
Then the branch of synchronized stationary
states bifurcates {\em subcritically} from
incoherence at $K^* =\infty$, provided
$\Omega_0>\Omega_0^{\infty}$. 
\end{itemize}

Summarizing, inertia favors the subcritical
character of bifurcations describing the
transition from incoherence to the partial
synchronized  state. In fact in the bimodal case,
the transition of incoherence to synchronization
will most likely occur as a subcritical
bifurcation as inertia increases. For a
continuous unimodal Lorentzian frequency
distribution, inertia may turn subcritical the
supercritical bifurcation to stationary
synchronized states, which is always found in the
masslesss Kuramoto model.

\section{High frequency limit}

\begin{figure}
\centerline{\hbox{\psfig{figure=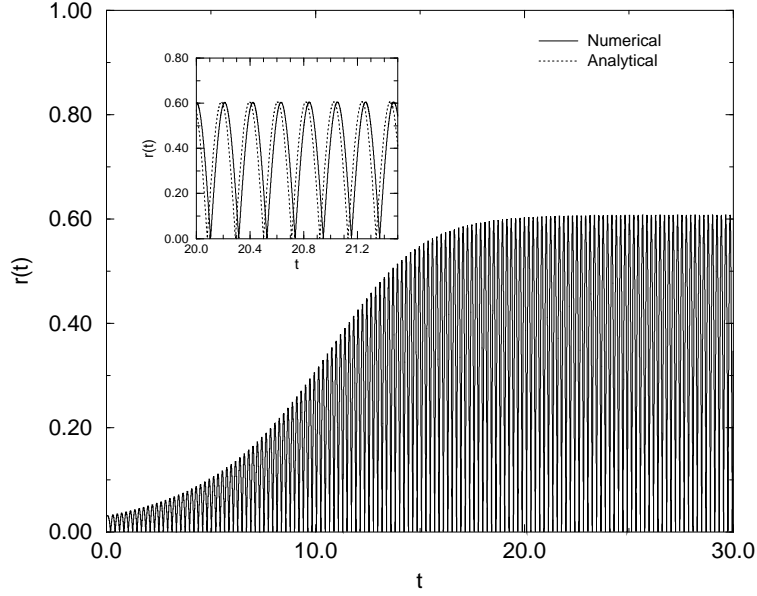,width=10.0cm}}}
\caption{Stable standing wave solution (SW) for discrete
bimodal frequency distribution. We have depicted the
evolution of the order parameter amplitude $r(t)$ for a 
large $\Omega_0=15$. Other parameter values are $m=0.1$,
$D=1$ and $K=5$. The inset shows a comparison between the
numerical solution and the leading-order asymptotic
approximation in the high-frequency limit. }
\label{sw1}
\end{figure}

\begin{figure}
\centerline{\hbox{\psfig{figure=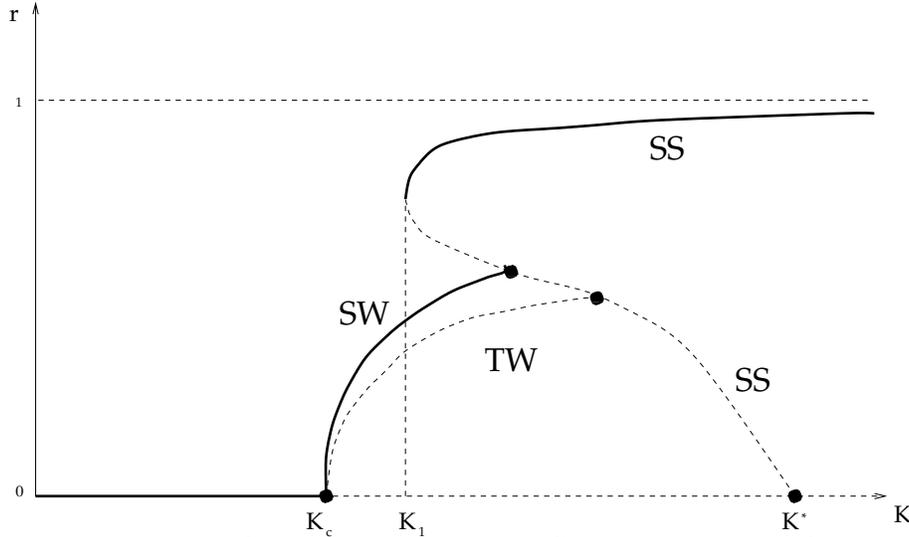,width=12.0cm}}}
\caption{Discrete bimodal frequency distribution: Schematic
global bifurcation diagram for positive values of $\alpha$
and $\beta$ (therefore the bifurcation of the stationary
synchronized solution (SS) from incoherence is subcritical).
Standing and traveling wave solutions are also shown.}
\label{bdiag}
\end{figure}

The high frequency limit of Eq.\ (\ref{fpe}) can
be analyzed  by means of a multiscale method. For
the Kuramoto model \cite{acebron2}, this method
lead to the result that the probability density
is (to leading order) a superposition of different
components corresponding to the different peaks
of the oscillator frequency distribution. To
apply this method to the present model, we
shall assume that the frequency distribution 
$g(\Omega)$ has $m$ maxima located at
$\Omega_0 \nu_l,l =1,....,m$, so that
$g(\Omega)d\Omega$ tends to the distribution
\begin{equation}
\Gamma(\nu)\equiv\sum_{l=1}^{m} \alpha_l
\delta(\nu-\nu_l)d\nu,
\end{equation}
as $\Omega_0\to\infty$. In order to simplify the
calculations, we change variables to a comoving
frame:
\begin{equation}
\beta=\theta-\Omega t\equiv \theta-\frac{\nu}{
\epsilon}t,
\end{equation}
and let $\omega'=\omega-\Omega$, where
\begin{equation}
\epsilon=\frac{1}{\Omega_0}\ll 1
\end{equation}
Then we obtain the following equations: 
\begin{eqnarray}
\frac{\partial \rho_j}{\partial t}=D
\frac{\partial^{2} \rho_j}{\partial
\omega'^{2}}-\frac{1}{m}\frac{\partial}{\partial
\omega'}(\rho_j U_j) -\omega'\frac{\partial
\rho_j}{\partial\beta},\\
U_j=-\omega'+\mbox{Im}\, K \left\{\sum_{l=1}^{m}
\alpha_l e^{i(\nu_l-\nu_j)t/\epsilon}
\right.\nonumber\\
\left. \times \int e^{i(\beta'-\beta)}
\rho(\beta',\omega',t,\nu_l;\epsilon) d\beta'
d\omega'\right\},\\
\int_{-\infty}^{\infty}\int_0^{2\pi}
\rho_j(\beta',t,\omega';\epsilon) d\beta'
d\omega'=1,
\end{eqnarray}
where
$\rho_j=\rho(\beta,t,\omega',\nu_j;\epsilon)$,
and $\rho\sim \sum \alpha_j\rho_j
\delta(\nu-\nu_j)$, \cite{acebron1}. We now
define fast and slow time scales,
$\tau=t/\epsilon$ and $t$, and make the Ansatz:
\begin{equation}
\rho=\sum_{n=0}^{2}
\rho^{(n)}(\beta,\omega',\tau,t,\nu)\epsilon^{n}+
O(\epsilon^3).
\end{equation}
Inserting this into the governing equations, we
obtain the following hierarchy:
\begin{eqnarray}
{\partial\rho_{j}^{(0)}\over\partial\tau } = 0,
\label{mm11}\\
{\partial\rho_{j}^{(1)}\over\partial\tau } =
- {\partial\rho_{j}^{(0)}\over \partial t} + D
{\partial^{2}\rho_{j}^{(0)} \over\partial
\omega'^{2}} +\frac{1}{m}{\partial\over \partial
\omega'} (\omega'\rho_j^{(0)})
\nonumber\\
- \omega' \frac{\partial\rho_j^{(0)}}{\partial
\beta} - \frac{K}{m} \frac{\partial}{\partial
\omega'} \left\{\rho_{j}^{(0)}\left[
\mbox{Im}\left(\alpha_{j} e^{-i \beta}
Z_{j}^{(0)} \right.  \right.\right.
\nonumber\\
\left.\left.\left. +  \sum_{l\neq j} \alpha_j \,
e^{i(\nu_{l} - \nu_j) \tau}\, e^{-i\beta}
Z_l^{(0)} \right) \right] \right\} \,,
\label{hierar}
\end{eqnarray}
where
\begin{equation}
Z_j^{(0)}(t) = \int_{-\infty}^{\infty}\int_{0}^{2\pi} e^{i\eta}
\rho_j^{(0)}(\eta,\omega,t,\nu_j)\,
d\eta\, d\omega.
\end{equation}
Elimination of secular terms yields the following
condition:
\begin{eqnarray}
&-& {\partial\rho_{j}^{(0)}\over \partial t} +
D {\partial^{2}\rho_{j}^{(0)} \over\partial\omega'^{2}} -\omega'
\frac{\partial \rho_j^{(0)}}{\partial \beta} \nonumber\\
&-& \frac{1}{m}
\frac{\partial} {\partial \omega'}\left\{\rho_{j}^{(0)}
\left[-\omega'+ K \alpha_{j} \,  \mbox{Im}\, (e^{-i \beta}
Z_{j}^{(0)}) \right]\right\}  = 0.
\end{eqnarray}
Notice that this equation corresponds to the
equation of a unimodal frequency distribution,
already studied in \cite{acebron2}. Its solution
evolves towards the following stationary state as
time evolves:
\begin{eqnarray}
\rho_j^{(0)}(\beta,\omega') &=& \sqrt{\frac{m}{
2\pi\,D}}e^{-(m/2D)
\omega'^2}\nonumber\\
&\times & \frac{e^{(K\,\alpha_j/D)R_j
\cos(\Psi_j-\beta)}}{\int_0^{2\pi}
e^{(K\,\alpha_j/D)R_j\cos(\Psi_j-\beta')}
d\beta'},\label{swsol}
\end{eqnarray}
where
\begin{eqnarray}
R_j e^{i\Psi_j} = \int_{-\infty}^{\infty}
\int_0^{2\pi} e^{i\eta}
\rho_j^{(0)}(\eta,\omega) d\eta d\omega \equiv
\lim_{t\to\infty} Z_j^{(0)}(t).
\end{eqnarray}
Incoherence of a component corresponds to $R_j=0$.
We know that as $K\alpha_j$ surpasses $2D$, a 
stable synchronized solution bifurcates from
incoherence. In the particular case of a symmetric
bimodal frequency distribution, the bifurcation
value is $K=4D$, independently of the inertia
$m$. This agrees with the results of the linear
stability analysis for $\Omega_0\to\infty$. All
the results previously obtained for the Kuramoto
model can be applied to the present model without
any modification \cite{acebron2}. Thus, both a
stable standing wave solution (SW) and an
unstable travelling wave solution (TW) bifurcate
supercritically from incoherence at $K=4D$
\cite{bps,acebron1}. In Figure \ref{sw1}
illustrates the comparison between the asymptotic
solution (\ref{swsol}) and the result of direct
numerical simulation for a large enough value
of the frequency $\Omega_0$. Finally, Fig.\
\ref{bdiag} shows the global bifurcation diagram
of the bimodal case for the case of positive
$\alpha$ and $\beta$. As explained in the
previous Section, the branch of synchronized
stationary states bifurcates {\em subcritically}
from incoherence at $K^* =\infty$, provided
$\Omega_0>\Omega_0^{\infty}$. Thus $K^*$ and the
end of the SW and TW branches in Fig.\
\ref{bdiag} should extend to infinity as
$\Omega_0 \to +\infty$. 

\section{Numerical results}
Four different numerical methods have been used.
Numerical simulations of the system of Langevin
equations (\ref{1}) were carried out for a large
number of oscillators (N=20 000), using an Euler
method.  A standard finite difference method was
used to solve numerically the Fokker-Planck
equation (\ref{fpe}), or the system of partial
differential  equations (\ref{systemc}). In
addition, we have used a simple spectral method,
which generalizes the one proposed in 
\cite{acebron2}. The idea is to solve a set of
ordinary differential equations for moments of
$\rho$:
\begin{eqnarray}
(x_i^j)_{k}:= \int_0^{2\pi}
c_i(\theta,\Omega_k,t)\cos [j(\psi-\theta)]\,d\theta,
\label{momentx}\\ 
(y_i^j)_{k}:= \int_0^{2\pi}
c_i(\theta,\Omega_k,t) \sin [j(\psi-\theta)]\,d\theta ,
\label{momenty}
\end{eqnarray}
\begin{eqnarray}
r=\int_0^{2\pi} \int_{-\infty}^{\infty}\int_{-\infty}^{\infty}
 \rho(\theta,\omega,\Omega,t)
\cos(\psi-\theta) \nonumber\\
\times g(\Omega)\, d\Omega\, d\omega\, d\theta\,.
\label{rnumerical}
\end{eqnarray}
The coefficient functions $c_i(\theta,\Omega,t)$
are the same as in the parabolic cylinder
expansion (\ref{expansion}). The integral in
(\ref{rnumerical}) will be approximated by a
suitable quadrature formula, as
\begin{eqnarray}
r\approx \sum_{q=1}^{Q}\alpha_q \int_0^{2\pi}
c_0(\Omega_q,\theta,t) \cos(\psi-\theta)\,d\theta
= \sum_{q=1}^{Q}\alpha_q(x_0^1)_{q} .
\end{eqnarray}
For instance, the Gauss-Laguerre quadrature has
been chosen case for the case of a Lorentzian
frequency distribution. The system of ordinary
differential equations is given by:
\begin{eqnarray}
(\dot{x}_i^j)_{k}=j\sqrt{i}\sqrt{\frac{D}{m}}
(y_{i-1}^j)_{k}+
\sqrt{i}\frac{1}{\sqrt{m\,D}}\Omega_k(x_{i-1}^j)_{k}
\nonumber\\
+ \sqrt{\frac{i}{m\,D}}\frac{K\,r}{2}
\left[(y_{i-1}^{j+1})_{k}-
(y_{i-1}^{j-1})_{k}\right]
-\frac{i}{m}(x_i^j)_{k}\nonumber\\
+\sqrt{i+1}\sqrt{D}{m}j
(y_{i+1}^j)_{k}-j\dot{\psi}(y_i^j)_{k},\\
(\dot{y}_i^j)_{k}=-j\sqrt{i}\sqrt{\frac{D}{m}}
(x_{i-1}^j)_{k}+\sqrt{i}\frac{1}{\sqrt{m\,D}}
\Omega_k(y_{i-1}^j)_{k}\nonumber\\
+\sqrt{\frac{i}{m\,D}}\frac{K\,r}{2}
\left[(x_{i-1}^{j-1})_{k}-
(x_{i-1}^{j+1})_{k}\right]
-\frac{i}{m}(y_i^j)_{k}\nonumber\\
-\sqrt{i+1}\sqrt{D}{m}j
(x_{i+1}^j)_{k}+j\dot{\psi}(x_i^j)_{k},\\
i=1,\ldots,N,\,\, j=1,\ldots,M, \nonumber
\end{eqnarray}
\begin{eqnarray}
(\dot{x}_0^j)_{k}=j\sqrt{\frac{D}{m}}
(y_{1}^j)_{k}-j\dot{\psi} (y_{0}^j)_{k},\\
(\dot{y}_0^j)_{k}=-j\sqrt{\frac{D}{m}}
(x_{1}^j)_{k}+j\dot{\psi} (x_{0}^j)_{k},\\
i=0,\,\, j=1,\ldots,M.  \nonumber
\end{eqnarray}

In order to numerically simulate this system, it
is necessary to truncate the hierarchy after a
reasonable number of modes $N$, and $M$. These
numbers will depend on the inertia $m$, the
spread, and the coupling strength. They should be
chosen large enough, so that the numerical results
do not depend on $N$ and $M$.

\begin{figure}
\centerline{\hbox{\psfig{figure=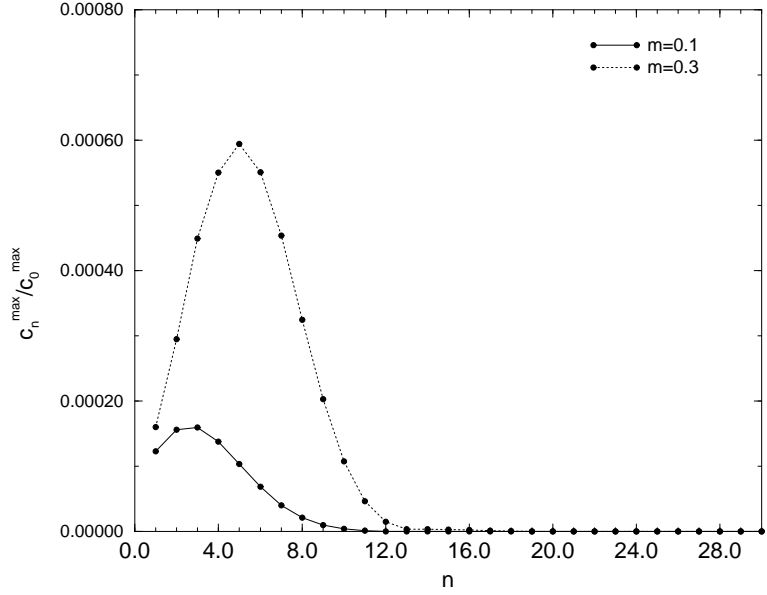,width=10.0cm}}}
\caption{Discrete bimodal frequency distribution: Ratio
$c_n^{max}/c_0^{max}$ (corresponding to the stationary
synchronized solution) as a function of $n$ for various mass
values and $\Omega_0=1$, $D=1$. The coefficients $c_n$ are
calculated numerically by integrating the time-dependent
equations and waiting until a stationary profile is reached.}
\label{nr1}
\end{figure}

\begin{figure}
\centerline{\hbox{\psfig{figure=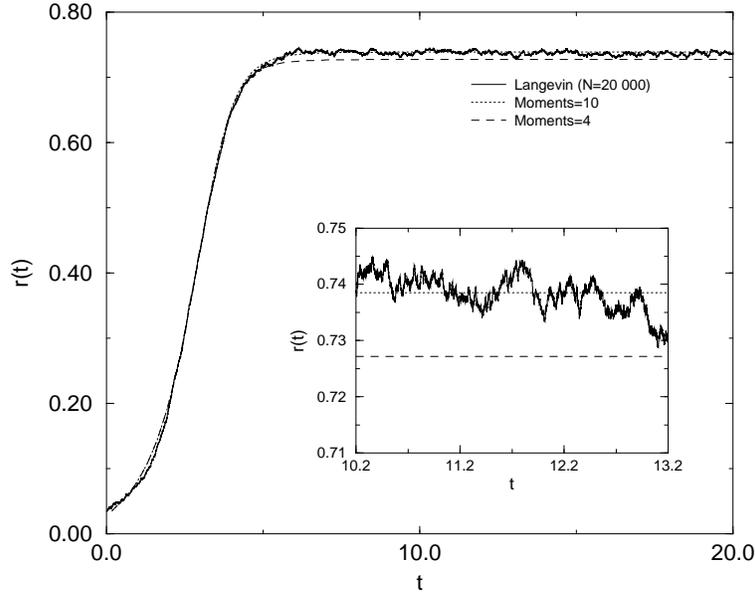,width=10.0cm}}}
\caption{Unimodal Lorentzian frequency
distribution: Comparison between the numerical
solution of a system of $N=20000$ Langevin
equations, and the numerical method proposed in
\S V containing moments of order 4 or 10.
Parameters are $m=0.2$, $D=1$, $\varepsilon=1$,
and $K=8$, and we have used $Q=15$ quadrature
nodes.}
\label{nr2}
\end{figure}

Fig.\ \ref{nr1} shows the ratio $c_n^{max}/ c_0^{max}$ as
a function of $n$, for the stationary solution and
for various mass values. $c_n^{max}$ is the
maximum value of the stationary coefficient
$c_n(\theta,\Omega_0=1)$. It has been obtained
from a finite-difference solution of Eq.\
(\ref{fpe}). Note that this ratio increases
as inertia does. Thus the truncation
approximation (\ref{sdis}) ceases to make sense
for larger values of inertia. Whether truncating
(at higher order) the moment equations
(\ref{momentx}) and (\ref{momenty}) yields good
numerical results, is tested in Fig.\ \ref{nr2}. This
figure shows how closely the previous system
(with moments of order 4 or 10) resembles the
direct solution of the Langevin equations. We
notice that the system containing moments of
order 10 is rather close to the solution
of the Langevin equations, but it does not
contain the fluctuations unavoidable in
stochastic methods. If we are interested in
solving the nonlinear Fokker-Planck equation, the
system of moment equations seems a good
alternative to solving Langevin equations for a
large number of oscillators.

\section{conclusions}
We have investigated synchronization properties of
a model of globally and nonlinearly coupled phase
oscillators, where the effects of white noise,
inertia and spread in the natural frequency
distribution are all considered. The linear
stability of the incoherent solution is rigorously
analyzed. Stationary and time-dependent solutions
of the standing and travelling wave type are
obtained by a variety of perturbative methods.
These include finding and approximately solving
mode-coupling equations for expansions of the
probability distribution in parabolic cylinder
functions, finding amplitude equations near
bifurcation points, and hierarchy-closure
assumptions. Numerical simulations of the
different equations and the original model
favorably agree with the different perturbative
results.

Inertia changes the stability boundaries of the
incoherent solution in a non trivial way. In the
case of an unimodal Lorentzian frequency
distribution, incoherence is stabilized, but the
effect of mass completely drops out if the
frequency spread vanishes. For a discrete bimodal
frequency distribution, both stability boundaries
and the character of the transition from
incoherence to synchronized states depend on the
values of the natural frequency $\Omega_0$ and on
inertia. The effect of inertia on the stationary
solution is dramatic in some cases. In general,
inertia hardens the synchronization transition:
it may render subcritical (hard) originally
supercritical (soft) transitions (in unimodal
Lorentzian frequency distributions), or it
increases the region in parameter space where the
transition is subcritical (discrete bimodal
frequency distributions). Analytical as well as
numerical calculations confirm these findings.

\section{Acknowledgments}
This work was supported, in part, by the European
Union TMR contract ERB FMBX-CT97-0157, by UNESCO
under contract UVO--ROSTE 875.629.9, by the
Spanish DGES grant PB98-0142-C04-C01, and by the
GNFM of the Italian C.N.R. All numerical
simulations were conducted at CASPUR--Rome.

\appendix
\section{Integrals of special functions and eigenvalue equation}
In (\ref{beta8}) there appear two integrals over $w$:
\begin{eqnarray}
A= \int_{-\infty}^{+\infty} e^{\left({w\over 2}  -
i\sqrt{mD}\right)^{2}} D_p \rho'_{0} dw,\nonumber\\
B= \int_{-\infty}^{+\infty}
e^{-\left({w\over 2} - i\sqrt{mD}\right)^{2}} D_{p}\, dw
\nonumber
\end{eqnarray}
The second integral is directly ${\cal A}_p(\sqrt{mD})$. The
first integral equals
\begin{eqnarray}
A= - {-m\over (2\pi)^{{3\over 2}} D}\,\int_{-\infty}^{+\infty}
(w - i2\sqrt{mD})\, e^{-\left({w\over 2} -
i\sqrt{mD}\right)^{2}} D_p(w) dw,\nonumber
\end{eqnarray}
because of (\ref{beta7}). An equivalent expression is
\begin{eqnarray}
A=  {2m\over (2\pi)^{{3\over 2}} D}\,\int_{-\infty}^{+\infty}
D_p(w)\, {\partial\over\partial w}\, e^{-\left({w\over 2}-
i\sqrt{mD}\right)^{2}}\,  dw.\nonumber
\end{eqnarray}
This can be written as
\begin{eqnarray}
A=  {2m\over (2\pi)^{{3\over 2}} D}\,\int_{-\infty}^{+\infty}
D_p(w)\, {\partial\over\partial w}\, e^{-{1\over 4}\left(w -
i2x\right)^{2}}\,  dw\nonumber\\
= {m\over i(2\pi)^{{3\over 2}} D}\,\int_{-\infty}^{+\infty}
D_p(w)\, {\partial\over\partial x}\, e^{-{1\over 4}\left(w -
i2x\right)^{2}}\,  dw\nonumber
\end{eqnarray}
provided we set $x=\sqrt{mD}$ after performing the calculations.
Thus we see that
\begin{eqnarray}
A= - {m\over i(2\pi)^{{3\over 2}} D}\, {\cal A}'_p(\sqrt{mD}).
\nonumber
\end{eqnarray}
This expression and the previous one for $B$ yield (\ref{beta9}).

Let us now obtain ${\cal A}_p(x)$ for integer $p$ and $x>0$. By using
that $D_p(x)$ is even  for even $p$ and odd for odd $p$, we can write
\begin{eqnarray}
{\cal A}_p(x) = e^{x^{2}}\,\int_{-\infty}^{+\infty}
e^{-{w^{2}\over 4} + iwx}\, D_p(w)\, dw\nonumber\\
= e^{x^{2}}\,\int_{-\infty}^{+\infty}
e^{-{w^{2}\over 4}} (\cos xw + i \sin xw)\, D_p(w)\, dw\nonumber\\
= 2 e^{x^{2}}\,\int_{0}^{+\infty} D_p(w)\,
e^{-{w^{2}\over 4}}\, f(xw)\, dw,\nonumber
\end{eqnarray}
where $f(xw) = \cos xw$ for $p$ even and $f(xw) = i \sin xw$ for $p$
odd. We can find (cf.\ \S 7.741 of \cite{gra}):
\begin{eqnarray}
\int_{0}^{+\infty} D_p(w)\, e^{-{w^{2}\over 4}}\, \cos xw\, dw
= (-1)^n \sqrt{{\pi\over 2}}\, e^{-{x^{2}\over 2}}\, x^{2n}\,,
\nonumber\\
\int_{0}^{+\infty} D_p(w)\, e^{-{w^{2}\over 4}}\, \sin
xw\, dw = (-1)^n \sqrt{{\pi\over 2}}\, e^{-{x^{2}\over 2}}\,
x^{2n+1}\,, \nonumber
\end{eqnarray}
provided $x>0$. From these formulas, we obtain
\begin{eqnarray}
{\cal A}_{2n}(x) = (-1)^n \sqrt{2\pi}\, e^{{x^{2}\over 2}}\, x^{2n}\,,
\nonumber\\
{\cal A}_{2n+1}(x) = i (-1)^n \sqrt{2\pi}\, e^{{x^{2}\over 2}}\,
x^{2n+1}\,. \nonumber
\end{eqnarray}
These two expressions are equivalent to (\ref{beta11}).

\section {Calculation of the coefficients
$\alpha$ and $\beta$}

We will now calculate coefficients $\alpha$ and
$\beta$ of Eq.\ (\ref{ralpha}).  The idea is
simply to expand the right hand side in the
definition of the order parameter,
\begin{equation}
r=\int_0^{2\pi} \int_{-\infty}^{\infty}
\int_{-\infty}^{\infty} 
\cos(\psi-\theta)
\rho(\theta,\omega,\Omega,t)\,g(\Omega)\,d\Omega\,d\omega\,d\theta.\label{order1}
\end{equation}
as a power series in $r$. To this end, we fix $r$
and $\psi$ in Eq.\ (\ref{fpe}), expand its
stationary solution in powers of $r$ and insert
the result in (\ref{order1}). As we do not have a
closed formula for the stationary solution of
(\ref{fpe}), we shall first derive a hierarchy of
equations for its coefficients in the series in
powers of $r$. Thus we have
\begin{eqnarray}
\rho(\theta,\omega,\Omega;r) = \sum_{n=0}^{\infty}
{G_{n}(\theta-\psi,\omega,\Omega)\over n!}\, r^n ,
\label{g}\\
r=\sum_{n=0}^{\infty} {r_{n}\over n!}\, r^n .
\label{expansionr}
\end{eqnarray}
Then
\begin{equation}
r_n=\int_0^{2\pi}\int_{-\infty}^{\infty}\int_{-\infty}^{\infty}
  \cos\theta
G_n(\theta,\omega,\Omega)g(\Omega)\,d\Omega\,d\omega\,d\theta,\label{r_nformula}
\end{equation}
where we have absorbed $\psi$ in the definition
of $G_n$. Comparison with eq. (\ref{ralpha}),
establishes that $r_0 = r_2 = 0$, $r_1=
K\alpha/2D$, and $r_3=K^3\beta$. From (\ref{g})
and (\ref{fpe}), we obtain
\begin{equation}
\frac{D}{m^2}\frac{\partial^2 G_n}{\partial
\omega^2}+\frac{1}{m}
\frac{\partial}{\partial \omega}
[(\omega-\Omega)G_n]-\omega
\frac{\partial G_n}{\partial \theta} =
-\frac{n}{m}K\sin \theta
\frac{\partial G_{n-1}}{\partial \omega},
\end{equation}
with $G_{-1}\equiv 0$. Since we are trying to
find solutions bifurcating from incoherence, we
should have $G_0= \rho_0(\omega,\Omega)$, i.e.\
the $\theta$-independent incoherent solution
(\ref{incoherent}). This directly confirms that
$r_0 = 0$.

The unknowns $G_n(\theta,\omega,\Omega)$ are
$2\pi$-periodic in $\theta$, so that we may
Fourier expand them as
\begin{equation}
G_n(\theta,\omega,\Omega)=\sum_{l=-\infty}^{
\infty} Z_n^l(\omega,\Omega) e^{i\,l\theta},
\end{equation}
where
\begin{equation}
Z_n^l(\omega,\Omega)=\frac{1}{2\pi}\int_0^{2\pi}
 e^{-i\,l\theta} G_n(\theta,\omega,\Omega)\,d\theta.
\label{znl}
\end{equation}
Here $Z_0^l \equiv 0$ for $l\neq 0$ because $G_0$
is the $\theta$-independent incoherent solution.
The condition that the probability density
function be real yields $\overline{G_{n}} =G_n$,
which in turn implies $Z_n^{-l} = \overline{
Z_{n}^{l}}$. Inserting these expressions into
(\ref{r_nformula}), we find
\begin{equation}
r_n=2\pi \int_{-\infty}^{\infty}\int_{-\infty}^{\infty} 
Re(Z_n^1)g(\Omega)\,d\Omega\,d\omega.
\label{rn}
\end{equation}
The unknowns $Z_n^l$ satisfy the following
hierarchy of equations:
\begin{equation}
\frac{D}{m^2}\frac{d^2 Z_n^l}{d \omega^2}+
\frac{1}{m} \frac{d}{d\omega}[(\omega-\Omega)
Z_n^l]-i\,l\omega Z_n^l=-\frac{n}{m}\frac{K}{2i}
\frac{d}{d\omega}[Z_{n-1}^{l-1}-Z_{n-1}^{l+1}].
\label{odeznl}
\end{equation}
The normalization condition for $\rho$ and the
incoherent solution together with (\ref{znl})
imply
\begin{equation}
\int_{-\infty}^{\infty} Z_n^0\, g(\omega) d\omega
= \delta_{n0} . \label{condzn0}
\end{equation}

 In order to obtain $\alpha$, and $\beta$, we
should show that $r_2=0$ and calculate $r_1$ and
$r_3$. We do this by means of (\ref{odeznl}),
with $n= 1,2,3$.\\

(a) Case n=1. Eq.\ (\ref{odeznl}) becomes
\begin{eqnarray}
\frac{D}{m^2}\frac{d^2 Z_1^1}{d \omega^2}+
\frac{1}{m} \frac{d}{d\omega}[(\omega- \Omega)
Z_1^1]-i\,\omega Z_1^1\nonumber\\
= -\frac{1}{m}\frac{K}{2i}
\frac{dZ_0^0}{d\omega} ,
\label{z11}
\end{eqnarray}
where $Z_0^0 = G_0$ is the incoherent solution
(\ref{incoherent}), and we have used $Z_0^2 =0$.
We can solve Eq.\ (\ref{z11}) by means of the
solution of Eq.\ (\ref{eqb}), in which we set
$\lambda=0$ and replace $b_1(\omega,\Omega)$ with
$Z_1^1(\omega,\Omega)$, and the right-hand side
with $i\frac{m\,K}{2D}dZ_{0}^{0}/d\omega$. Then
we obtain
\begin{eqnarray}
Z_1^1(\omega;\Omega) &=& -{i K\over 2m}\,
e^{-\frac{m}{4D}(\omega-\Omega)^2}\nonumber\\
&\times & \sum_{p=0}^{\infty} {\int_{-\infty}^{\infty} e^{\left({w\over
2}  - i\sqrt{mD}\right)^{2}} D_p
\frac{dZ_0^{0}}{d\omega} dw\over \sqrt{2\pi} p!
\left({p\over m} + i\Omega + D\right)}\, D_p(w) , \label{z11sol}
\end{eqnarray}
The previous analogy allows us to calculate $r_1$
from (\ref{rn}) and Equations (\ref{beta8}),
(\ref{beta10}) and (\ref{beta11}) (with
$\lambda=0$). The result is
\begin{eqnarray}
r_1=\frac{K D}{2}e^{m\,D}
\left[\int_{-\infty}^{\infty}
\frac{1}{D^2+\Omega^2}g(\Omega)\,d\Omega \right.\nonumber\\
\left.+\sum_{p=1}^{\infty}
 \frac{(-m\,D)^{p}(1+\frac{p}{m\,D})^{2}}{p!}
\right.\nonumber\\
\left.\times \int_{-\infty}^{\infty} \frac{1}{(D+\frac{p}{m})^2+\Omega^2}
g(\Omega)\,d\Omega\right].
\end{eqnarray}

(b) Case n=2. We now show that $r_2=0$.\\
In order to analyze $r_2$, it is necessary to
consider $Z_2^1$, which is the solution to the
following equation:
\begin{equation}
\frac{D}{m^2}\frac{d^2 Z_2^1}{d \omega^2}
+\frac{1}{m} \frac{d}{d\omega}[(\omega-\Omega)
Z_2^1] -i\,\omega Z_2^1=-\frac{2}{m}\frac{K}{2i}
\frac{d}{d\omega}[Z_{1}^{0}-Z_{1}^{2}].
\end{equation}
We shall show that $Z_{1}^{0}=Z_{1}^{2}=0$. If
this is so, the resulting homogeneous equation
can be transformed to the parabolic cylinder
equation with a quadratic potential having
complex coefficients. Its solution cannot decay
to zero both as $\omega\to\pm\infty$ unless it is
identically zero. The reason $Z_{1}^{0}$ and
$Z_{1}^{2}$ are zero is similar. The equation for
$Z_{1}^{0}$ is homogeneous ($Z_0^l =0$ for
$l\neq 0$), and it has a solution
$C(\Omega) e^{-m(\omega-\Omega)^2/(2D)}$. The
normalization condition (\ref{condzn0}) then
implies $C\equiv 0$.  $Z_1^2$ again obeys a
homogeneous equation which can be transformed
into the parabolic cylinder equation with a
complex potential. The only solution which
decays to zero as $\omega\to\pm\infty$ is
again $Z_{1}^{2}=0$.

(c) Case n=3.\\
The equations for $Z_3^1$ are
\begin{eqnarray}
\frac{D}{m^2}\frac{d^2 Z_3^1}{d \omega^2}+
\frac{1}{m}
\frac{d}{d\omega}[(\omega-\Omega)Z_3^1]-i\,\omega
Z_3^1=-\frac{3}{m}\frac{K}{2i}
\frac{d}{d\omega}[Z_2^0-Z_2^2],\nonumber\\
\frac{D}{m^2}\frac{d^2 Z_2^2}{d \omega^2}+
\frac{1}{m}
\frac{d}{d\omega}[(\omega-\Omega)Z_2^2]-2 i\,\omega
Z_2^2=-\frac{2}{m}\frac{K}{2i}
\frac{d}{d\omega}Z_1^1,\nonumber\\
\frac{D}{m^2}\frac{d^2 Z_2^0}{d \omega^2}+
\frac{1}{m}
\frac{d}{d\omega}[(\omega-\Omega)Z_2^0]=
-\frac{2}{m}\frac{K}{2i}
\frac{d}{d\omega}[Z_1^{-1}-Z_1^1]=\frac{2K}{m}\,
Im\frac{d}{d\omega}Z_1^1,\nonumber\\
\frac{D}{m^2}\frac{d^2 Z_1^1}{d \omega^2}+
\frac{1}{m}
\frac{d}{d\omega}[(\omega-\Omega)Z_1^1]-i\,\omega
Z_1^1=-\frac{1}{m}\frac{K}{2i}
\frac{d}{d\omega}Z_0^0.\label{z3sys}
\end{eqnarray}
This system of equations has to be solved
numerically in order to obtain the coefficient
$\beta$.

%%%%%%%%%%%%%%%%%% Linear Stability Figures %%%%%%%%%%%%%

%\begin{figure}
%\centerline{\hbox{\psfig{figure=fig9.ps,width=12.0cm}}}
%\centerline{\hbox{\psfig{figure=fig91.ps,width=7.0cm}}
%\hbox{\psfig{figure=fig92.ps,width=7.0cm}}}
%\centerline{\hbox{\psfig{figure=fig93.ps,width=7.0cm}}
%\hbox{\psfig{figure=fig94.ps,width=7.0cm}}}
%\caption{Lorentzian unimodal frequency
%distribution: Stability diagram of incoherence in
%the parameter space ($m$,$K$) for $\varepsilon
%=1$, $D=1$.  The amplitude of the order
%parameter as a function of time is numerically
%calculated and displayed at the points marked by:
%(1) $K=6$, $m=0.4$; (2) $K=6$, $m=0.8$; (3)
%$K=20$, $m=4.25$; (4) $K=20$, $m=4.75$. }
%\label{fig9}
%\end{figure}

\newpage
%\begin{figure}
%%\centerline{\hbox{\psfig{figure=fig6.ps,width=12.0cm}}}
%\caption{Discrete bimodal frequency distribution:
%Real and imaginary parts of the eigenvalue
%$\lambda$ as functions of $m$, for $K=4.6$,
%$\Omega_0=0.8$ and $D=1$.}
%\label{fig6}
%\end{figure}

%%%%%%%%%%%%%%%%%%%%%%%%%%%%%%%%%%%%%%%%%%%%%%%%%%%%%%%%%%%%%%%%%

%%%%%%%%%%%%%%%%%% Stationary Solution Figures %%%%%%%%%%%%%%%%%% 

%%%%%%%%%%%%%%%%%%%%%%%%%%%%%%%%%%%%%%%%%%%%%%%%%%%%%%%%%%%%%%%%%%%%%%

%%%%%%%%%%%%%%%%%%% High Frequency Limit Figures %%%%%%%%%%%%%%%%%%%%%

%%%%%%%%%%%%%%%%%%%%%%%%%%%%%%%%%%%%%%%%%%%%%%%%%%%%%%%%%%%%%%%%%%%%%

%%%%%%%%%%%%%%%%%%%%% Numerical Results %%%%%%%%%%%%%%%%%%%%%%%%%%%%

%%%%%%%%%%%%%%%%

%\end{multicols}
\end{document}